\documentclass[fleqn,usenatbib]{mnras}

\usepackage{newtxtext,newtxmath}

\usepackage[T1]{fontenc}
\usepackage{ae,aecompl}

\usepackage[authoryear]{natbib}
\usepackage{psfrag,color}
\usepackage[latin1]{inputenc}
\usepackage{graphicx}	
\usepackage{amsmath}	

\newcommand{\Ne}{{$n_{\rm e}$}}
\newcommand{\Te}{{$T_{\rm e}$}}
\newcommand{\TO}{{$T_{\rm e}$[\ion{O}{iii}]}}
\newcommand{\To}{{$T_{\rm e}$[\ion{O}{ii}]}}
\newcommand{\TS}{{$T_{\rm e}$[\ion{S}{iii}]}}
\newcommand{\TN}{{$T_{\rm e}$[\ion{N}{ii}]}}

\newcommand\W{{$\lambda$}}
\def\i{\,{\sc i}}

\newcommand{\hb}{H$\beta$} 
\newcommand{\ha}{H$\alpha$}

\newcommand{\oiii}{O\thinspace{\sc iii}} 
 
\newcommand{\nii}{N\thinspace{\sc ii}}

\newcommand{\oii}{O\thinspace{\sc ii}}

\newcommand{\ciii}{C\thinspace{\sc iii}}

\usepackage{orcidlink}

\title[Chemical evolution of CNO]{The JWST EXCELS survey: direct estimates of C, N, and O abundances in two relatively metal-rich galaxies at $\mathbf{z\simeq5}$}

\author[K. Z. Arellano-C\'ordova]{
K. Z. Arellano-C\'ordova \orcidlink{0000-0002-2644-3518} $^{1}$\thanks{E-mail: ziboney@gmail.com and k.arellano@ed.ac.uk (KZAC)},
F. Cullen \orcidlink{0000-0002-3736-476X}$^{1}$,
A. C. Carnall \orcidlink{0000-0002-1482-5818}$^{1}$,
D. Scholte \orcidlink{0000-0002-6867-1244}$^{1}$,
\and \
T. M. Stanton \orcidlink{0000-0002-0827-9769}$^{1}$,
C. Kobayashi \orcidlink{0000-0002-4343-0487}$^{2}$,
Z. Martinez \orcidlink{0000-0000-0000-0000}$^{3}$,
D. A. Berg \orcidlink{0000-0002-4153-053X}$^{3}$,
L. Barrufet $^{1}$, \and \
R. Begley \orcidlink{0000-0003-0629-8074},$^{1}$, 
C. T. Donnan \orcidlink{0000-0002-7622-0208}$^{1}$, 
J. S. Dunlop $^{1}$,
M. L. Hamadouche \orcidlink{0000-0001-6763-5551}$^{4}$,
D. J. McLeod \orcidlink{0000-0003-4368-3326}$^{1}$, \and\
R. J. McLure $^{1}$, 
K. Rowlands \orcidlink{0000-0001-7883-8434}$^{5,6}$,
A. E. Shapley$^{7}$
\\
$^{1}$ Institute for Astronomy, University of Edinburgh, Royal Observatory, Edinburgh EH9 3HJ, UK \\
$^{2}$ Centre for Astrophysics Research, Department of Physics, Astronomy and Mathematics, University of Hertfordshire, Hatfield, AL10 9AB, UK\\
$^{3}$ Department of Astronomy, The University of Texas at Austin, 2515 Speedway, Stop C1400, Austin, TX 78712, USA\\
$^{4}$Department of Astronomy, University of Massachusetts, Amherst, MA 01003, USA;\\
$^{5}$ AURA for ESA, Space Telescope Science Institute, 3700 San Martin Drive, Baltimore, MD 21218, USA\\
$^{6}$ William H. Miller III Department of Physics and Astronomy, Johns Hopkins University, Baltimore, MD 21218, USA\\
$^{7}$Department of Physics \& Astronomy, University of California, 430 Portola Plaza, Los Angeles CA 90095, USA\\
}

\date{Accepted XXX. Received YYY; in original form ZZZ}

\pubyear{2020}

\begin{document}
\label{firstpage}
\pagerange{\pageref{firstpage}--\pageref{lastpage}}
\maketitle

\begin{abstract}
We present a spectroscopic analysis of two star-forming galaxies at $z\simeq5$ observed with {\it JWST}/NIRSpec as part of the EXCELS survey. 
The detection of the \ion{C}{iii}]~\W\W1906,09, [\oii] \W\W3726,29, [\oiii] \W\W4363,5007, and [\nii] \W6584  emission lines enables an investigation of the $\mathrm{C/O}$, $\mathrm{N/O}$, and $\mathrm{C/N}$ abundance ratios using the temperature-sensitive method.
The galaxies have stellar masses of ${\mathrm{log}(M_{\star}/\mathrm{M}_{\odot}) = 8.09^{+\, 0.24}_{-0.15}}$ and ${\mathrm{log}(M_{\star}/\mathrm{M}_{\odot}) = 8.02^{+\, 0.06}_{-0.08}}$ with metallicities of $Z \simeq 0.2 \, \rm{Z_{\odot}}$ and $Z \simeq 0.3 \, \rm{Z_{\odot}}$.
These metallicities are somewhat higher than is typical for other $z\gtrsim 5$ galaxies with similar stellar mass and are comparable  to $z \simeq 0$ analogues. 
Both galaxies display evidence for elevated N/O ratios with respect to the typical star-forming galaxies at $z\simeq0$, with ${\mathrm{log(N/O)} = -1.07^{+\,0.17}_{-0.13}}$ and ${\mathrm{log(N/O)} = -0.86^{+\,0.15}_{-0.11}}$ respectively. 
In contrast, we find low C abundances, with ${\mathrm{log(C/O)}=-0.82\pm0.22}$ and ${\mathrm{log(C/O)}=-1.02\pm0.22}$, consistent with the predicted yields of core-collapse supernovae. 
Following the trend observed in other high-redshift sources, we find that the $\mathrm{C/N}$ ratios are lower at fixed $\mathrm{O/H}$ compared to the majority of local galaxies.
Via a comparison to detailed chemical evolution models, we find that a standard or bottom-heavy IMF can explain the observed abundance ratios where the N-enrichment comes from intermediate mass ($\simeq 4-7 \, \mathrm{M}_{\odot}$) stars.
Our results demonstrate that robust measurements of CNO abundances with \emph{JWST} can reveal unique enrichment pathways in galaxies as a function of both metallicity and redshift.
\end{abstract} 



\section{Introduction}
\label{sec:intro}

Carbon (C), nitrogen (N), and oxygen (O) are important tracers of the chemical enrichment of the interstellar medium (ISM) in star-forming regions. 
Since these elements are formed on different timescales, the relative abundances of C and N in relation to O (i.e, $\mathrm{C/O}$ and $\mathrm{N/O)}$ are crucial for tracing the star-formation history and chemical evolution of galaxies.

Nitrogen is produced in both massive stars ($> 10 \, \mathrm{M}_{\odot}$) and intermediate-mass stars ($\simeq 4 - 7 \, \mathrm{M}_{\odot}$). 
At high metallicity, the majority of N production arises from intermediate mass stars during the asymptotic giant branch (AGB) phase, while at low metallicities, massive stars eject N stellar winds on short timescales ($\sim$25 Myr) \citep[e.g.,][]{henry00, kobayashi11, vincenzo16, kobayashi20}. 
Carbon is created in triple-$\alpha$ reactions within massive stars and also within low-mass AGB stars ($\simeq 1 - 4 \, \mathrm{M}_{\odot}$). 
The ejection of oxygen (and most other $\alpha$-elements) occurs at core-collapse supernovae (CCSNe) of massive stars (on $\sim$40 Myr timescale) \citep[e.g.,][]{kobayashi20, arellanocordova24}. 
Oxygen is used as a proxy of the total metallicity of the ISM, and is typically parametrised with the number ratios as 12+log(O/H).

While oxygen and nitrogen abundances can be analysed via the rest-frame optical spectra of photoionised objects \citep[e.g.,][]{perez-montero09, croxall16, arellanocordova20, berg19b, berg16, berg21a, izotov23}, the faintness of the optical \ion{C}{ii} \W4267 recombination lines (RLs) makes them extremely difficult to detect across a wide range of physical conditions \citep[e.g.,][]{esteban05, skillman20}. 
However, the \ion{C}{iii}]~\W 1906 and [\ion{C}{iii}]~\W1909~\footnote{We represent the \ion{C}{iii}] and [\ion{C}{iii}] lines as \ion{C}{iii}] \W\W1906,09 in this work.} collisionally excited lines (CELs) are among the brightest features in the ultraviolet (UV) regime, enabling analyses of the $\mathrm{C/O}$ ratio in galaxies across a wide range of redshifts~\citep[e.g.,][]{garnett90, berg16, berg19b, pena-guerrero17, arellanocordova22b}. 
 
Observations with the James Webb Space Telescope ({\it JWST}) have recently provided crucial information on additional ionisation states of N that can be observed in the rest-frame UV, facilitating studies of N enrichment in the very early Universe \citep[][]{Bunker23}. The \ion{N}{IV}]~\W\W$1483,86$ and \ion{N}{iii}]~\W\W1750-54 UV lines are now frequently used to derive the N/O in objects at $z \gtrsim 6$ \citep{Cameron23, marqueschaves23, isobe23b, schaerer24, topping24b, topping24a, curti24b, ji24}. 
Interestingly, chemical abundance studies using the UV N lines have revealed extremely high $\mathrm{N/O}$ enrichment in some early sources (e.g., GN-z11 at $z=10.6$; \citealp{Bunker23}; \citealp{Cameron23}).
Such high $\mathrm{N/O}$ ratios are puzzling, since the nucleosynthesis of N and O are expected to follow similar pathways at low metallicity \cite[e.g.,][]{berg16, berg19a, vincenzo16}.  

Several scenarios have been proposed to explain these elevated N abundance in star-forming galaxies (SFGs) at the highest redshifts. 
Possible explanations include winds from very massive stars \citep[$\gtrsim 100 \, \mathrm{M}_{\odot}$;][]{vink23} and/or super massive stars~\citep[$\simeq 10^{4} \, M_{\odot}$;][]{charbonnel23, watanabe24}, precursors of globular clusters~\citep[e.g.,][]{senchyna23}, the presence of active galactic nuclei (AGN; \citealp{maiolino24}) or shocks and Wolf-Rayet (WR) stars \citep{flury24}. 
Additionally, \citet{kobayashi24} used chemical evolution models to reproduce the $\mathrm{C/O}$ and $\mathrm{N/O}$ abundance ratios of GN-z11, concluding that galaxies with two bursts of star-formation followed by a quiescent phase of $\simeq$ 100 Myr can successfully reproduce the observed abundance patterns. 
\citet{kobayashi24} suggest that the quiescent phase can be related to outflows, while the first burst of star-formation can be triggered by pre-enriched gas from external galaxies.

Although most recent high-redshift N abundance determinations have been derived mainly with UV N lines, studies of SFGs at intermediate redshifts ($z\simeq2-3$) have used the optical [\ion{N}{ii}]~\W6584 line \citep{sanders23a, rogers24, welch24b}. 
These studies also typically return a high value of $\mathrm{N/O}$ in comparison with local SFGs, albeit less extreme than for the $z>6$ sources. 
The reason for these elevated $\mathrm{N/O}$ ratios has been attributed to pristine gas inflows that can dilute the metallicity at fixed N/O \citep{amorin10}, the presence of WR stars enriching the ISM \citep{riverathorsen24, welch24b} or the details of the star-formation history \citep[e.g.,][]{perez-montero09, sanders23a}. 
However, \citet{rogers24} recently reported a $\mathrm{N/O}$ ratio for a galaxy at $z \simeq 3$ consistent with local galaxies (i.e., no evidence for excess N enrichment); a similar result was reported by \citet{berg18} from a study of the rest-frame UV spectrum of a lensed galaxy at $z \simeq 2$.

On the other hand, the first estimates of the 
$\mathrm{C/O}$ ratio in SFGs at $z>6$  \citep{arellanocordova22b, jones23} have been broadly consistent with the results of low-metallicity SFGs at $z\simeq0$ (i.e., $\mathrm{12+log(O/H)} < 8.0$; \citealp{berg12}).
For example, \citet{jones23} analysed the $\mathrm{C/O}$ ratio of a galaxy at $z=6.23$ using chemical evolution models, concluding that the production of C is consistent with the expectations of pure CCSNe enrichment at low metallicities. 
Similarly low $\mathrm{C/O}$ ratio ratios have been found in galaxies with super-solar N abundances at $z>6$ \citep[e.g.,][]{marqueschaves23, topping24a, schaerer24,curti24b}, indicating that while C is consistent with standard nucleosynthesis, an additional source of enrichment seems to be affecting the chemical abundance of N. 

However, although most studies report low $\mathrm{C/O}$ abundance ratios, evidence of enhanced C enrichment has also been found in a small number of high-redshift sources. 
For example, \citet{Hsiao24b} analysed the spectrum of a lensed galaxy at $z=10.17$ (MACS0647-JD) and derived a value of $\mathrm{log(C/O)}=-0.44$, which is somewhat higher than other $z>6$ estimates (typically between $\mathrm{log(C/O)}=-0.8$ and $-1.1$; e.g., \citealp{jones23, topping24b, topping24a}).
These authors concluded that this high value might be due to the high metallicity derived for this object ($\mathrm{12+log(O/H)=7.79}$), or the presence of exotic stellar populations \citep{charbonnel23, watanabe24}. 
In addition, while \citet{castellano24} (GHZ2; $z=12.34$) reported a consistent $\mathrm{C/O}$ ratio compared to local low-metallicity galaxies, \citet{deugenio24} (Gs-z12; $z=12.5$) find a super-solar C abundance that they explain by appealing to the predicted yields of very metal-poor stars. 
However, an important caveat related to the studies of \citet{castellano24} and \citet{deugenio24} is that they have to assume a value for the electron temperature (\Te) of the gas when deriving abundances. 
Ultimately, a direct determination of \Te\ is needed to provide a robust determination of the chemical composition of the gas \citep[e.g.,][]{peimbert17}. 

A review of the current literature therefore suggests that a coherent picture of the enrichment of C and N in galaxies across all cosmic epochs has not yet emerged.
Significant systematic offsets and scatter seem to exist, possibly related to differences in the star-formation histories of galaxies, different enrichment mechanisms, variations in the stellar initial mass function (IMF), or observational systematics \citep[e.g.,][]{kobayashi20, curti24b, Yan17, Bekki23}. 
To make further progress, additional samples of galaxies with robust estimates of C, N and O abundances are clearly needed.

In this context, we present an analysis of the ionised gas and physical properties of two SFGs at $z\simeq5$ observed with JWST/NIRSpec.
These SFGs are part of the Early eXtragalactic Continuum and Emission Line Science survey \citep[EXCELS,][]{carnall24}. 
Due to the specific redshifts of these sources, they are the only two galaxies from EXCELS for which we can characterise the chemical abundance patterns of C, N, and O.
Crucially, in both sources we detect the [\oiii] \W4363 auroral line which probes the gas electron temperature, enabling robust abundance estimates using the direct method.
Currently, only four other sources with similar direct estimates of C, N, and O exist in the literature at $z \gtrsim 5$ \citep{topping24b, topping24a, curti24b}. Our analysis therefore represents a significant increase in direct CNO estimates at early cosmic epochs.
By combining the NIRSpec spectra with deep JWST PRIMER photometry, we derive the physical properties of these two sources to gain further insight into what drives their chemical enrichment. 

This paper is structured as follows. In Section~\ref{sec:Observations}, we describe the data reduction of the EXCELS galaxies, and we also describe the additional samples of galaxies across a range of redshifts that we use for comparison. 
In Section~\ref{sec:methodology}, we detail our methodology for the spectral energy distribution (SED) fitting, emission-line flux measurements, and the determination of chemical abundances; we also compare the stellar mass and the star-formation rate as a function of the stellar mass, and investigate the ionisation source via BPT diagrams.  
In Section~\ref{sec:CNO_results}, we present our main results on the C, N, and O abundance patterns by analysing the $\mathrm{N/O-}$, $\mathrm{C/O-}$, and $\mathrm{C/N-O/H}$ abundance diagrams. 
Finally, we summarise our results and present our conclusions in Section~\ref{sec:conclusion}. 

In this paper we assume a standard cosmological model with $H_0=70$\,km s$^{-1}$ Mpc$^{-1}$, $\Omega_m=0.3$ and $\Omega_{\Lambda}=0.7$, and use the Solar metallicity scale of \citet{asplund21}, where ${\mathrm{12 + log(O/H)_\odot} = 8.69}$,  ${\mathrm{log(C/O)_\odot} = -0.23}$, and ${\mathrm{log(N/O)_\odot} = -0.86}$.

\begin{table*}
    \centering
    \caption{A compilation of star-forming galaxies at $z\ge2$ with measurements of the metallicity (O/H), N/O and C/O abundance ratios derived using the \Te-sensitive method. 
    Columns 1 and 2 give the galaxy ID and its redshift, while columns $3-6$ give the metallicity and the C/O and N/O abundance ratios.  
    For N/O, we have separated the results based on whether UV N lines or optical N lines were used to derive the abundance ratio (columns 5 and 6). 
    Column 7 provides the references for the original studies.}
    \begin{tabular}{lcrrrrl}
    \hline
  Galaxy   &  $z$ & 12+log(O/H) &  log(C/O)  &  log(N/O)$_{\rm {opt}}$  & log(N/O)$_{\rm {UV}}$  & Reference \\

\hline
 SL2S J0217-0513 & 1.84 &  $\geq7.50$  & $-0.81\pm0.09$ & $-$ & $-1.48\pm0.46$ & \citet{berg18} \\

ID$\_$19985	 & 2.19 &  $7.89\pm0.20$ & $-$ & $-0.69\pm0.10$ & $-$ & \citet{sanders22} \\
ID$\_$20062	 & 2.19 &  $8.24\pm0.27$ & $-$ & $-0.78\pm0.14$ & $-$ & \citet{sanders22} \\
Sunburst Arc	  & 2.37 & $7.97\pm0.05$ & $-$ & $-0.65^{+0.16}_{-0.25}$ & $-$ & \cite{welch24b} \\
Q2343-D40	 & 2.96  & $8.07\pm0.06$ & $-$ & $-1.37\pm0.21$ & $-$ & \citet{rogers24} \\
J0332-3557	 & 3.77 &  $8.26\pm0.06$ & $-1.02\pm0.20$ & $-$ & $-$ & \citet{citro24} \\
RXCJ2248-ID &  6.11  & $7.43\pm0.17$ & $-0.83\pm0.11$ &  $-0.60\pm0.15$$^{*}$ & $-0.39\pm0.11$ & \citet{topping24a} \\
A1703-zd6 & 7.04 & $7.47\pm0.19$ & $-0.70\pm0.18$ &  -- &  $-0.60\pm0.30$ & \cite{topping24b}\\
CEERS$\_$1019 & 8.63 & $7.70\pm0.18$ & $-0.75\pm0.11$ & -- & $-0.18\pm0.28$ & \citet{marqueschaves23}  \\
GLASS$\_$15008 &  6.23 & $7.39\pm0.23$ &  $-1.01\pm0.12$ & -- & -- &\cite{jones23}\\
GLASS$\_$15008 & 6.23 & $7.65^{+0.14}_{-0.08}$ & $-1.08^{+0.06}_{-0.14}$ & -- & $-0.40^{+0.05}_{-0.07}$ & \cite{isobe23a}\\
ERO$S\_$04590 & 8.49 & $7.12\pm0.12$ &  $-0.83\pm0.38$ & -- & -- &\citet{arellanocordova22b}\\
GN-z9p4  & 9.38  &  $7.37\pm0.15$  & $>-1.18$ &  -- & $-0.59\pm0.24$ & \citet{schaerer24}\\
GS-z9-0  & 9.43  &  $7.40\pm0.09$  & $-0.93\pm0.24$ &  -- & $-0.90\pm0.24$ & \citet{curti24b}\\
MACS0647-JD  & 10.17  &  $7.79\pm0.09$  & $-0.44^{+0.06}_{-0.07}$ &  - & $-$ & \citet{Hsiao24b}\\
\hline
EXCELS-121806 & 5.225 & $7.97^{+0.05}_{-0.04}$ & $-1.02\pm0.22 $ & $-0.87^{+0.14}_{-0.11}$   & -- & This Work \\
EXCELS-70684 & 5.255  &  $8.21^{+0.08}_{-0.05}$ & $-0.82\pm0.22 $ & $-1.07^{+0.17}_{-0.12}$   & -- & This Work \\
\hline
\end{tabular}
 \label{tab:high-z_sample}
 \begin{description}
     \item[] * Optical N/O abundance that we have derived in this study using the flux measurements and the physical conditions reported in \citet{topping24a}.
 \end{description}
\end{table*}

\section{Data}
\label{sec:Observations}
The two galaxies presented here, EXCELS-70864 and EXCELS-121806, were observed as part of the EXCELS Survey (GO 3543; PIs: Carnall, Cullen; \citealp{carnall24}).
The EXCELS survey consists of four NIRSpec MSA pointings within the PRIMER UDS field (GO 1837; PI: Dunlop); a detailed description of the survey can be found in \citet{carnall24}.
Here, we briefly summarize the observing strategy and describe the data reduction. 

\subsection{Observations and data reduction}

Each of the four NIRSpec/MSA pointings in EXCELS were observed with the medium-resolution ($\mathrm{R} \simeq 1000$) G140M/F100LP, G235M/F170LP and G395M/F290LP gratings, and observations were carried out using a 3 shutter slitlet and 3-point dither pattern.
For each pointing, the total integration times in each grating were $\simeq 4$ hours in G140M and G395M, and $\simeq 5.5$ hours in G235M, using the NRSIRS2 readout pattern.
As described in \citet{carnall24}, separate MSA configurations were specified for each grating and targets were observed in various combinations of the three gratings.
The two objects presented here were observed in all three NIRSpec gratings\footnote{ Because independent MSA configurations were used for each grating, there are small variations in the slit position with respect to the galaxy centroid between the gratings. However, these shifts are negligible for the two objects presented here.}.
At the redshift of these galaxies, the observations covered rest-frame wavelengths $\lambda_{\rm rest} \simeq 1600 - 8500${\AA} which provides access to a number of key rest-frame UV emission-lines such as \ion{He}{ii}~\W1640 and \ion{C}{iii}]~\W\W1906,1909 and several rest-frame optical lines such as [\ion{O}{ii}]~\W3726,3727, \hb, [\ion{O}{iii}]~\W4363, \ha, and [\ion{N}{ii}]~\W\W6548,6584. 

The raw data were reduced using v1.15.1 of the JWST reduction pipeline\footnote{https://github.com/spacetelescope/jwst}.
We adopted the default level 1 configuration except for turning on advanced snowball rejection and using the CRDS$\_$CTX = jwst$\_$1258.pmap version of the JWST Calibration Reference Data System (CRDS) files.
We then ran the level 2 pipeline steps assuming the default configuration.
Finally, the level 3 pipeline was used to combine the 2D spectra that were then used for science analysis.
We then performed our own custom 1D optimal extraction of the 2D spectra \citep{horne86}, setting the extraction centroid as the flux-weighted mean position of the object within the NIRSpec/MSA slitlet \citep[see][]{carnall24}.

\begin{figure*}
\begin{center}
    \includegraphics[width=0.90\textwidth, trim=35mm 5mm 35mm 5, clip=yes]{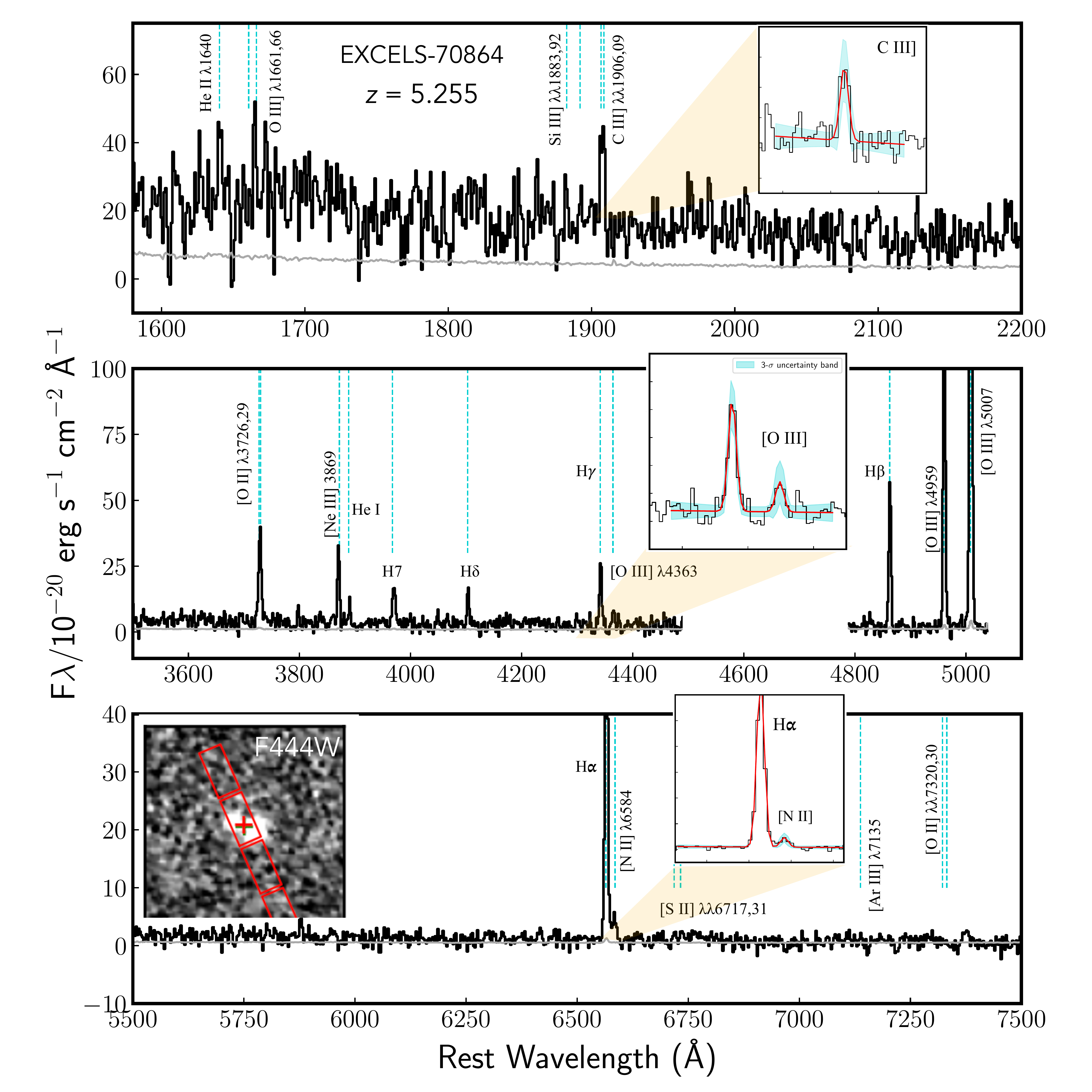}
        \caption{Rest-frame UV and optical NIRSpec spectra of EXCELS-70864 at $z=5.255$. 
        The black and grey line show the spectrum and error spectrum respectively. 
        We highlight the detection of some UV and optical features such as \ion{C}{iii}] \W1906,1909, [\ion{O}{ii}] \W\W3726,29, [\ion{Ne}{iii}] \W3869, [\ion{O}{iii}] \W4363, [\ion{N}{ii}] \W6584, and [\ion{S}{ii}] \W\W6717,31, which are used to determine the physical conditions, chemical abundances and physical properties of EXCELS-70864. 
        The insert figures show the emission-line fitting of \ion{C}{iii}] \W1906,1909, [\ion{O}{iii}] \W4363 and [\ion{N}{ii}]~\W6584 which are essential to calculate the C and N abundances using the \Te\-sensitive method. 
        The emission line fits are shown in red with the uncertainties are shown by the light blue shaded region.
        The bottom panel shows the F444W imaging and the slit position of the G235M grating. The crosses indicate the galaxy center (green) and the slit center (red)} \citep[see also][]{carnall24}.
\label{fig:spec_1}
\end{center}
\end{figure*}

\begin{figure*}
\begin{center}
    \includegraphics[width=0.90\textwidth, trim=35mm 5mm 35mm 5, clip=yes]{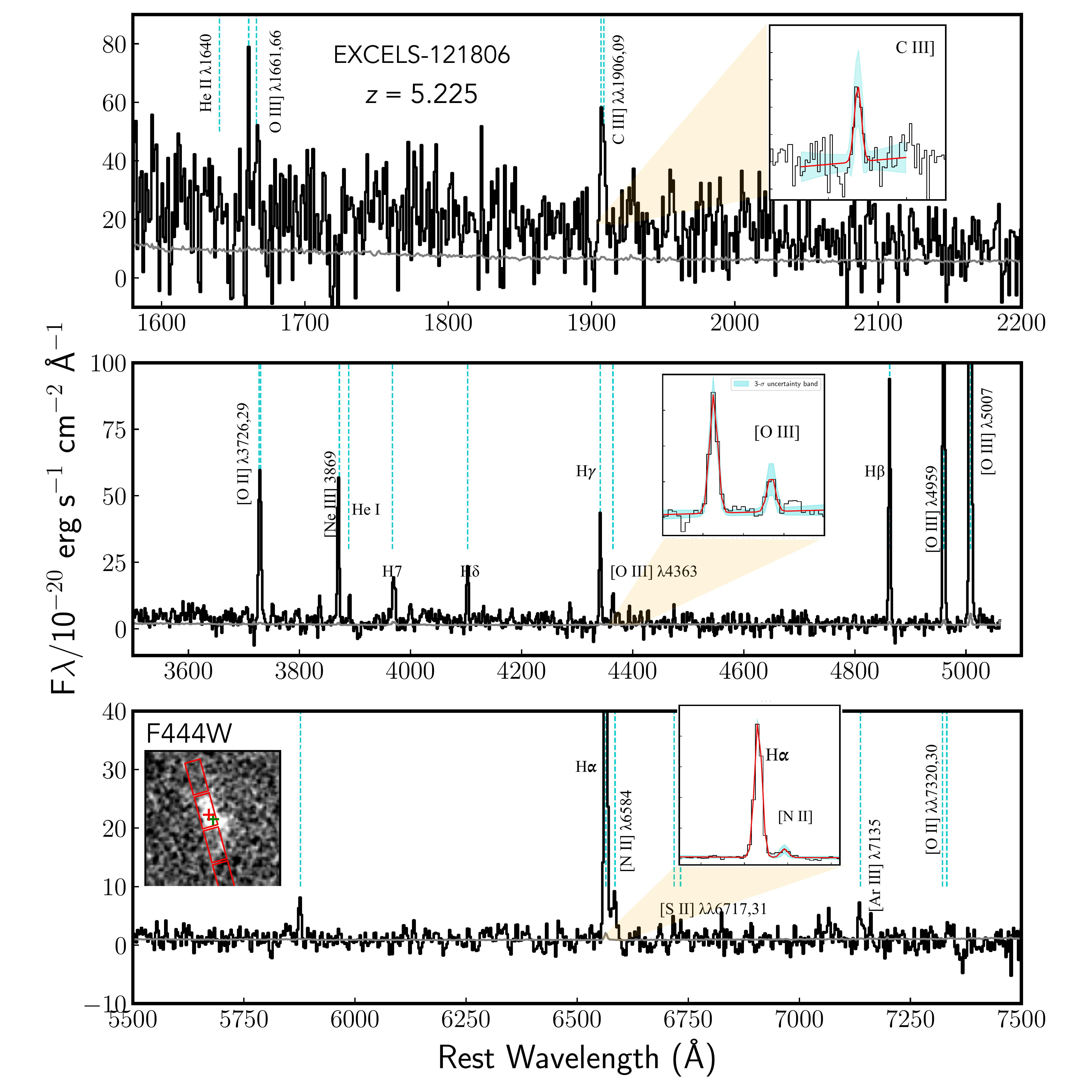}
        \caption{Rest-frame UV and optical  NIRSpec spectra of  EXCELS-121806 at $z=5.255$. For more details see Fig.\ref{fig:spec_1}.} 
\label{fig:spec_2}
\end{center}
\end{figure*}

We performed flux calibration by first matching the flux level between the different gratings.
To do this, we calculated the median flux in overlapping wavelength regions and scaled the G140M and G395M spectra to the overlapping regions of the G235M spectra.
To achieve an absolute flux calibration, we then scaled the full spectra to the overlapping broadband photometry.
Each galaxy in the EXCELS sample benefits from deep HST and JWST/NIRCam imaging spanning the wavelength range $0.4 - 5 \mu \mathrm{m}$ in 11 photometric bands: F435W, F606W, F814W, F090W, F115W, F150W, F200W, F277W, F356W, F410M and F444W.
The broadband fluxes are extracted from the PSF-homogenised images using $0.5$ arcsec diameter apertures and corrected to total using the F356W FLUX$\_$AUTO values measured by \textsc{SourceExtractor} \citep{bertin_sextractor}.
We integrated each spectrum through the appropriate overlapping filters and scaled them to the corresponding photometry.
The final correction as a function of wavelength is derived via a linear interpolation between the bands.
Figs.~\ref{fig:spec_1} and~\ref{fig:spec_2} show the final rest-frame UV-to-optical spectra for EXCELS-70864 and EXCELS-121806. 
From the multiple emission lines visible in these figures, we measure spectroscopic redshifts of $z=5.255$ and $z=5.225$ respectively.

\subsection{Additional star-forming galaxy samples}
\label{sec:add-sampple}

To supplement our analysis, we have also gathered a sample of star-forming galaxies across a range of cosmic epochs with the aim of comparing their properties and chemical patterns with our EXCELS galaxies at $z\simeq5$. 
For this literature selection, we have only considered galaxies that have robust C, N and O abundance determinations using the \Te-sensitive method. 

This comparison sample includes dwarf SFGs at $z\simeq0$ from \citet{berg12}, \citet{berg16}, \citet{berg19a}, \citet{izotov23}, and \citet{flury22a}. 
In particular, we have incorporated galaxies from the COS Legacy Archival Spectroscopic SurveY (\citealp{Berg22, james22, arellanocordova24}; CLASSY), which spans a wide range in ISM conditions and physical properties (e.g., stellar mass, SFR, and ionisation parameter). 
The CLASSY galaxies span a metallicity range of ${\mathrm{12+log(O/H)}=7.9-8.8}$ and represent our primary local analogue sample. We have taken the results of C and N from Martinez et al. in prep. and Arellano-C\'ordova et al. in prep, respectively. 
Additionally, we have also included the $z\simeq0$ Lyman continuum emitting (LCE) galaxies from \citet{izotov23}; as discussed in \citet{izotov23}, these galaxies appear to show a distinct CNO abundance pattern compared to non-LCE galaxies \citep[e.g.,][]{berg12} which may help to interpret the abundance patterns being observed at higher redshifts.

Along with the local sample, we have also compiled measurements of the physical properties and direct CNO chemical abundances for a sample of galaxies in the redshift range $z=2-10$ taken from a number of literature sources \citep{berg18, sanders23a, llerena23, topping24a, morishita24, marqueschaves23, rogers24, schaerer24, citro24, hu24, isobe23b, curti24b,topping24b,Hsiao24b, shapley24}. 
In total, our comparison sample comprises measurements of C, N, and O abundances with metallicities ranging between $\mathrm{12+log(O/H) = 7.12}$ to $8.8$. 
We note that we have used abundance patterns directly from the original papers and not attempted to recalculate them in this work.
We have also compiled the galaxy properties, such as stellar mass and SFR, as reported by the original authors when these values are available; where necessary, we have converted the values to a \citet{kroupa01} IMF for consistency with our analysis.
In Table~\ref{tab:high-z_sample}, we list the chemical abundance determinations for this $z\gtrsim2$ literature sample.


\section{Chemical abundances and physical properties}
\label{sec:methodology}

In this section, we present our analysis of the JWST data for the two galaxies at $z \simeq 5$.  
We first describe the measurement of emission-line fluxes from the NIRSpec data. 
These flux measurements then allow us to determine the reddening due to dust, the ionisation source, star-formation rate, physical conditions of the ISM (\Te\ and \Ne), and the chemical abundances.
We finally describe how the stellar mass of these galaxies was estimated from fitting to the emission-line corrected broadband photometry.

\subsection{Emission-Line Measurements}
The emission-line fluxes are measured following the methodology of \citet{Scholte25} and \cite{Stanton25}. 
Here, we include a short summary of that procedure. All the emission-lines are fitted using Gaussian line profiles after continuum subtraction using a common intrinsic line width and line velocity. The line amplitude is freely fitted for each emission-line.
The continuum flux is estimated using a running mean of the 16$^{\rm th}$ to 84$^{\rm th}$ percentile flux values within a top hat function with a rest-wavelength width of 350 \AA. In Sec.~\ref{sec:galaxy_properties}, we provide a brief discussion on continuum subtraction fitting based on SED fitting. Details of the procedure can be found in \citet{Scholte25}. For the \ion{C}{iii}]~\W\W1907,09 and [\ion{O}{ii}]~\W\W3727 doublets,  we have modelled the flux as the  sum of the two Gaussian components. The intrinsic line widths of the individual emission lines are tied to the other emission lines. Total line widths are then also convolved with the wavelength dependent resolution of the grating.

To re-estimate the flux uncertainties we use the continuum-subtracted residuals by implying a multiplication factor to the pipeline flux uncertainties such that $\tilde{\sigma}_{F} = \frac{1}{2}(R_{84} - R_{16})$, where $\tilde{\sigma}_{F}$ is the median flux uncertainty and $R_{16}$, $R_{84}$ are the 16$^{\rm th}$, 84$^{\rm th}$ percentiles of the residuals. 
The emission-line flux uncertainties are calculated using a line profile weighted average of the pixel flux uncertainties \citep[e.g.,][]{moustakas23}. 
As discussed in \citet{Scholte25}, we find an excess scatter (i.e., after accounting for the statistical uncertainties) in line flux measurements of the same emission lines in multiple gratings of $\simeq8$ per cent.
Throughout this paper, this additional calibration uncertainty is included in any line ratio measured using lines measured in different gratings.
The inset panels in Figs.~\ref{fig:spec_1} and~\ref{fig:spec_2} shows some examples of the fitting for \ion{C}{iii}] \W\W1906,09,  the \Te-sensitive [\ion{O}{iii}]~\W4363 line, and [\ion{N}{ii}]~\W6584 for the two galaxies in this analysis (these are the three crucial lines required to simultaneously measure C, N and O abundances).  
In Table~\ref{tab:fluxes}, we present the fluxes and corresponding uncertainties of the emission-lines analyzed in this paper.

\begin{figure}
\begin{center}
    \includegraphics[width=0.35\textwidth, trim=0mm 0mm 0mm 0mm, clip=yes]{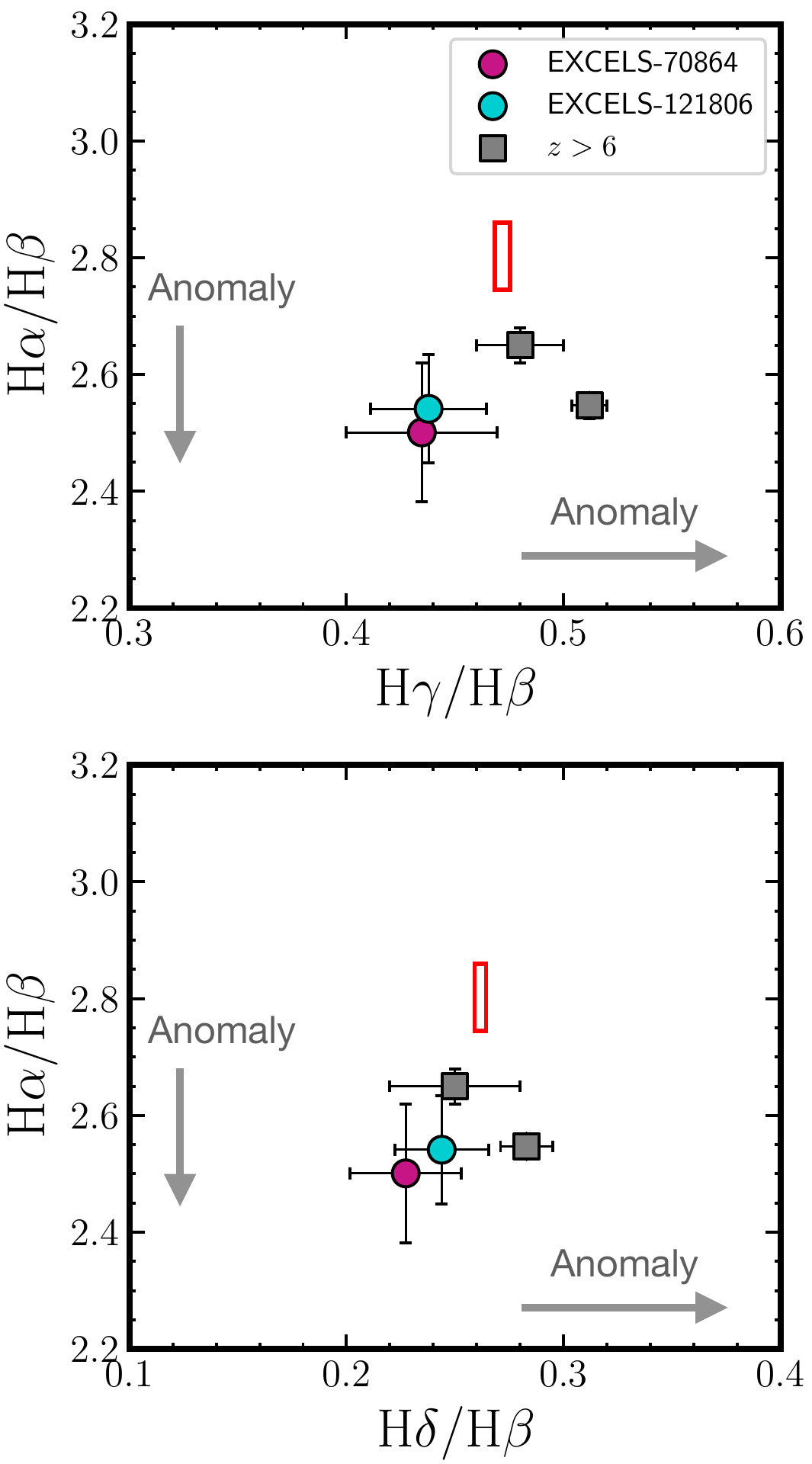}
        \caption{The observed Balmer ratios of the EXCELS galaxies at $z\simeq5$. 
        The red rectangles represent the theoretical Balmer ratios under the conditions of ${T_e = 10 \, 000 \, \rm{K} - 20 \, 000 \, \rm{K}}$ and $n_e = 300 \, \rm{cm}^{-3}$. 
        {\it Top panel}: H$\alpha$/H$\beta$ versus H$\gamma$/H$\beta$. 
        Both galaxies display H$\alpha$/H$\beta$ ratios below the Case B value while the H$\gamma$/H$\beta$ ratios are consistent with the Case B value or a small amount of nebular reddening. 
        {\it Bottom}: H$\alpha$/H$\beta$ versus H$\delta$/H$\beta$. 
        Again the EXCELS galaxies H$\delta$/H$\beta$ ratios consistent with the theoretical value within the uncertainties. 
        The gray squares represent the measurements of galaxies at $z\sim6$ taken from \citet{topping24a} and \citet{Cameron23}. 
        The arrows indicate the direction in which the Balmer ratios become anomalous with respect to the Case B theoretical values (i.e., not consistent with the dust-free Case B ratios).
        Consistent with other results in the literature, we find that both galaxies display sub-Case B H$\alpha$/H$\beta$ ratios.
        However, all Balmer line ratios are consistent with the zero-reddening Case B ratios within the $2\sigma$ uncertainties.} 
\label{fig:balmer}
\end{center}
\end{figure}

\begin{table}
    \centering
    \caption{Emission-line fluxes for the two EXCELS galaxies at $z\simeq5$.}
    \begin{tabular}{lrr}
    \hline
\multicolumn{1}{c}{} & \multicolumn{1}{c}{70864} & \multicolumn{1}{c}{121806} \\
\multicolumn{1}{l}{Line} & \multicolumn{1}{c}{Flux} & \multicolumn{1}{c}{Flux} \\
\multicolumn{1}{l}{} & \multicolumn{1}{l}{[10$^{-19}$ erg s$^{-1}$ cm$^{-2}$]} & \multicolumn{1}{l}{[10$^{-19}$ erg s$^{-1}$ cm$^{-2}$]} \\
 \hline
 \ion{He}{ii}~\W1640  &  $8.18\pm2.36$ & $<{ 2.96}$\\
 \ion{O}{iii}]~\W1666 & $<{6.79}$ & $9.44\pm3.10$\\ 
 \ion{C}{iii}]~\W\W1907,09$^{\star}$ & $11.55\pm2.07$ & $16.06\pm3.21$ \\
 $[\ion{O}{ii}]$~\W\W3727 & $21.90\pm1.30$ & $34.10\pm1.50$\\
 $[\ion{Ne}{iii}]$~\W3869 & $14.44\pm0.78$ & $23.58\pm1.03$ \\
 H$\delta$ & 6.42$\pm$0.69 & $9.63\pm0.08$\\
 H$\gamma$ & 12.28$\pm$0.88 & $17.29\pm0.93$\\
$[\ion{O}{iii}]$~\W4363 & $2.84\pm0.69$ & $5.43\pm0.84$ \\
H$\beta$ (G235M)$\dagger$&  $28.26\pm1.01$ & $39.49\pm1.15$\\
$[\ion{O}{iii}]$~\W4959 & $73.18\pm1.45$ & $94.32\pm1.46$\\
$[\ion{O}{iii}]$~\W5007 & $220.94\pm2.26$ & $276.74\pm2.30$\\
H$\beta$ (G395M)$\dagger$ & $25.64\pm1.13$ & $40.54\pm1.36$\\
H$\alpha$ & $64.13\pm1.13$ & $103.03\pm1.42$\\
$[\ion{N}{ii}]$~\W6584 & $2.77\pm0.56$ & $6.78\pm0.72$ \\
$[\ion{S}{ii}]$~\W6717 & $<{1.00}$ & $2.4\pm0.75$\\
$[\ion{S}{ii}]$~\W6731 &  $<{ 1.14}$ & $1.96\pm0.72$\\
$[\ion{Ar}{iii}]$~\W7135 & $<{ 0.74}$ & $4.77\pm0.90$ \\
 \hline
\end{tabular}
 \label{tab:fluxes}
\begin{description}
\item[]Notes: $^{\star}$ \ion{C}{iii}]~\W1907 + [\ion{C}{iii}]~\W1909. $\dagger$ H$\beta$ (G235M) and H$\beta$ (G395M) correspond to the flux measured from each grating. 
\end{description}
\end{table}

\subsection{Nebular dust reddening}\label{subsec:reddening}

We first compared the observed and theoretical values of the Balmer line ratios to assess the magnitude of dust reddening in EXCELS-70864 and EXCELS-121806 \citep{osterbrock89, osterbrock06}. 
For both galaxies, we have used the following Balmer ratios: $\mathrm{H}\delta/\mathrm{H}\beta$, $\mathrm{H}\gamma/\mathrm{H}\beta$, and $\mathrm{H}\alpha/\mathrm{H}\beta$. 
Although higher-order Balmer lines are also detected in both spectra (e.g., H$\epsilon$, H$\zeta$, and H$\eta$), we decided to omit these lines due to the contamination from nearby emission-lines and large uncertainties on their fluxes.  
We assumed Case B recombination corresponding to theoretical values of $\mathrm{H}\gamma/\mathrm{H}\beta = 2.78$,  $\mathrm{H}\gamma/\mathrm{H}\beta = 0.47$ and $\mathrm{H}\delta/\mathrm{H}\beta = 0.26$ which we calculated using the \texttt{PyNeb} package \citep{luridiana15} for $T_{\rm e} = 15,000 \, \rm{K}$ and $n_{\rm e} = 300 \, \rm{cm}^{-3}$. 

In Fig.~\ref{fig:balmer}, we report the observed Balmer ratios compared to the range of expected theoretical values computed for temperatures between $T_{\rm e}= 10,000 - 20,000 \, \rm{K}$ (red rectangles). 
For both galaxies, we do not find strong evidence for inconsistencies with the theoretical ratios within the uncertainties.
In both cases, there is evidence for minor amounts of dust reddening in their $\mathrm{H}\gamma/\mathrm{H}\beta$ and $\mathrm{H}\delta/\mathrm{H}\beta$ ratios, albeit in both cases the ratios are consistent with $E(B-V)=0$.
Interestingly, however, both galaxies show a value of $\mathrm{H}\alpha/\mathrm{H}\beta$ \emph{below} the theoretical ratio.
These anomalous ratios that fall systematically below the expected value are becoming relatively common in JWST spectra at high-redshifts \citep[e.g.,][]{Yanagisawa24}. 
To highlight this, we have also added in Fig.~\ref{fig:balmer} two other $z > 6$ galaxies which show anomalous $\mathrm{H}\alpha/\mathrm{H}\beta$ ratios from \citet{topping24a} (RXCJ2248-ID) and \citet{Cameron23} (GS-NDG-9422).
We note that low values of $\mathrm{H}\alpha/\mathrm{H}\beta$ have also been reported in galaxies at low-redshift \citep{kewley05, scarlata+24}. 

\begin{figure*}
\begin{center}
    \includegraphics[width=0.85\textwidth, trim=5mm 0mm 0mm 0mm, clip=yes]{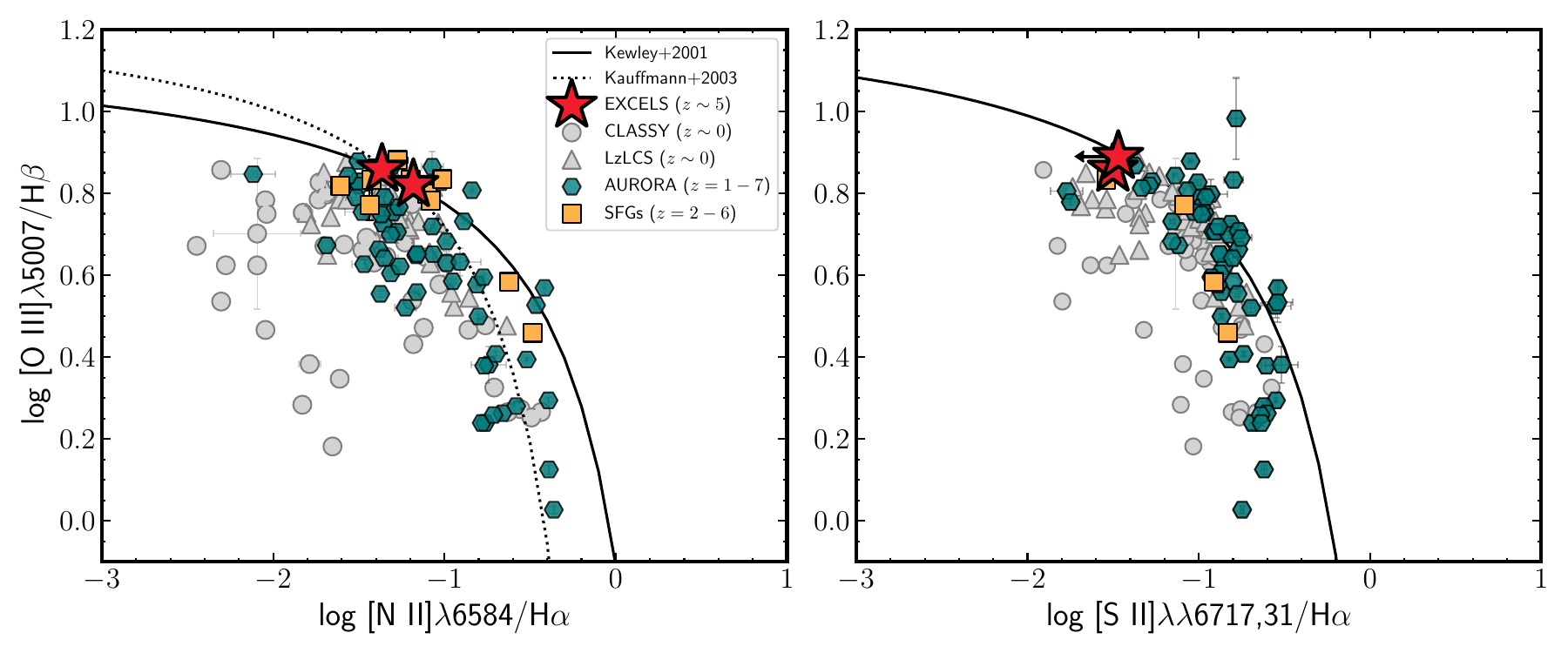} 
    \caption{The [\ion{N}{ii}] (left) and  [\ion{S}{ii}] (right) BPT diagrams for the EXCELS galaxies and a selection of literature samples. For EXCELS-70864, we have determined an upper limit for [\ion{S}{ii}]/H$\alpha$.
    The solid line shows the theoretical maximum starburst separation of \citet{kewley01}, while the dashed line illustrates the boundary between star-formation and AGN from \citet{kauffmann03}. 
    The orange squares show a sample of galaxies at high-redshift with [\ion{N}{ii}] and [\ion{S}{ii}] detections from \citet{sanders23a} ($z\simeq2$),  \citet{sanders24} ($z \simeq 2-4$). \citet[][]{rogers24} ($z =2.96$), \citet{welch24b} ($z =2.37$), and \citet[][]{topping24a} ($z = 6.11$). 
    The turquoise hexagons show data from the AURORA survey at $z=1-7$ \citep{shapley24}. 
    The gray symbols represent the CLASSY survey \citep{Berg22, mingozzi24} and SFGs from the Lyman Continuum Survey (LzLCS; \citealp{flury22a}). 
    The two EXCELS galaxies display similar line ratios to the other $z > 3$ galaxies, which are also consistent with the local comparison samples.
    There is no obvious evidence for AGN ionisation based on these optical emission line ratios.}
\label{fig:BPT}
\end{center}
\end{figure*}

Investigation of these anomalous values of $\mathrm{H}\alpha/\mathrm{H}\beta$ have been conducted in \cite{Yanagisawa24} and \citet{McClymont24} for galaxies at $z>6$ and by \cite{scarlata+24} for galaxies at $z\simeq0$. 
For example, \citet{McClymont24} analyzed the cause of anomalous $\mathrm{H}\alpha/\mathrm{H}\beta$ and $\mathrm{H}\gamma/\mathrm{H}\beta$ ratios using 19 galaxies from JADES. 
With the aid of photoionisation models, these authors conclude that a density-bounded nebula is one of the possible explanations for the anomalous ratios.
It is unclear whether EXCELS-70864 and EXCELS-121806 are consistent with density-bounded nebulae, although we note the [\ion{O}{iii}]~\W5007/[\ion{O}{ii}]~\W3726,3727 ratios are not particularly extreme (Table~\ref{tab:fluxes}). 

Since all ratios are consistent within $2\sigma$ of their Case B values, we have assumed that the anomalous Balmer ratios are most likely due to statistical scatter in our case.
From the $\mathrm{H}\gamma/\mathrm{H}\beta$ and $\mathrm{H}\delta/\mathrm{H}\beta$ ratios, we estimate ${E(B-V) = 0.19\pm0.11}$ for EXCELS-70864 and ${E(B-V) = 0.13\pm0.08}$ for EXCELS-121806. 
As both of these values are consistent with ${E(B-V)=0}$ within the uncertainties, and the $\mathrm{H}\alpha/\mathrm{H}\beta$ ratios also indicate zero/negligible dust attenuation, in the following analysis, we have used the observed fluxes without dust correction. 
Reassuringly, we find that in the case of zero reddening, the physical conditions derived for EXCELS-121806 using the optical [\ion{O}{iii}]~\W4363 line and UV [\ion{O}{iii}]~\W1666 line are fully consistent (see Table~\ref{tab:physical conditions}).
We therefore argue that the assumption of zero reddening is an appropriate for our analysis.
In Section ~\ref{sec:CNO_results}, we discuss the impact of dust uncertainties on the $\mathrm{C/O}$ and $\mathrm{N/O}$ abundance ratios.  

\subsection{Evidence for AGN ionisation?}

We investigated whether the observed line ratios are compatible with pure star-formation or whether there is evidence for AGN ionisation in either of our sources.
emission line ratio diagnostics can help us to discern how the gas in galaxies is being ionised (e.g., star-formation, AGN or shocks; \citealp{baldwin81, allen08, kewley01,kauffmann03, mingozzi24}). 
In Fig.~\ref{fig:BPT}, we show the [\ion{N}{ii}]/H$\alpha$ (left) and [\ion{S}{ii}]/H$\alpha$ (right) BPT diagrams for our two EXCELS galaxies; we also include the additional sample of local and high-redshift galaxies for comparison.
It can be seen from Fig.~\ref{fig:BPT} that both of the EXCELS galaxies are located close to the limit of the region of pure star-formation defined by the \citet{kewley01} and \citet{kauffmann03} demarcation lines (dotted and solid lines, respectively). 
The same is true for most of the other galaxies at $z>2$ (squares), and indicates very high ionisation objects. 
We also show the [\ion{S}{ii}]-BPT diagram for one of our galaxies, EXCELS-121806, which has a robust detection of the [\ion{S}{ii}] \W\W6717,31 doublet (see right panel of Fig.~\ref{fig:BPT}), finding a consistent picture in which the line ratios fall close to the star-formation limit. For EXCELS-70864, we used the upper limits of the [\ion{S}{ii}] $\lambda\lambda$6717, 31 fluxes to estimate the [\ion{S}{ii}]/H$\alpha$ ratio.
Overall, we find no strong evidence to suggest that the gas in either galaxy is being ionised by an AGN. 
Moreover, the spectra of these two galaxies do not show evidence for broad or multiple Gaussian components. 

It can be seen from Fig.~\ref{fig:BPT} that the high-redshift samples are systematically offset from the local analogue sample; this is a well-studied problem \citep{steidel14, shapley15, shapley24} that can be attributed, at least in part, to the $\alpha$-enhanced abundance ratios that appear to be ubiquitous in the $z>2$ star-forming population \citep[e.g.][]{steidel16, cullen19, topping20, cullen21, stanton24}.
At fixed $\alpha$-element abundance, high-redshift galaxies are expected to have systematically lower Fe abundances and therefore harder ionising spectra.
These line ratio diagrams therefore offer the first piece of evidence that our $z\simeq5$ galaxies likely differ from the local analogues in terms of their ionisation state and chemical abundances.

 The rest-frame UV lines are another alternative way to distinguish between star-formation and AGN driven ionisation in galaxies \citep{mingozzi24, topping24b, topping24a, curti24b}. 
 In addition to the UV carbon lines, we also have measurements of \ion{He}{ii} \W1640 (S/N $\ge$ 3) for EXCELS-70864.
 However, we are missing the detection of the \ion{O}{iii}] \W\W1661,66 doublet, which makes it difficult to analyze any UV-diagnostic \citep[e.g.,][]{mingozzi24, flury24}.
 In future work, we plan to characterize the ionisation source for the full EXCELS sample in greater detail.

\begin{table}
    \centering
    \caption{The physical properties and chemical abundances derived for the two $z\simeq5$ EXCELS galaxies.}
    \begin{tabular}{lrr}
    \hline
\multicolumn{3}{c}{EXCELS} \\
  Galaxy Properties                      &  70864  & 121806 \\
\hline
log M$_{\star}$ (M$_\odot$) &  {$8.09^{+0.24}_{-0.15}$} & { $ 8.02^{+0.06}_{-0.08}$}\\
log(SFR) (H$\alpha$) [M$_\odot$ yr$^{-1}$] & $0.79\pm0.01$ & $1.00\pm0.01$\\
log(SFR) (SED) [M$_\odot$ yr$^{-1}$] & { $0.97^{+0.09}_{-0.07}$} & { $1.02\pm0.07$}\\
\Ne[\ion{S}{ii}] [cm$^{-3}$] & -- & { $ 600^{+920}_{-400}$} \\
 \Te[\ion{O}{iii}] [K]  & $12500^{+1100}_{-1300}$&  $14900\pm1100$ \\
 \Te[\ion{O}{iii}] (UV)$^{\star}$ [K]  & $-$ &  $15300\pm1250$ \\
 12+log(O/H) & $8.21^{+0.08}_{-0.05}$ & $7.97^{+0.05}_{-0.04}$\\
 log(C/O) & $-0.82\pm0.22$ & $-1.02\pm0.22$\\
 log(C/O) (UV)$^{\star}$ & $-$ & $-1.07\pm0.43$\\

 log(N/O) & $-1.07^{+0.17}_{-0.13}$ &  $-0.86^{+0.15}_{-0.11}$  \\
 log(Ne/O) & $-0.70\pm0.16$ &  $-0.63\pm0.11$  \\
 log(Ar/O) & -- & $-2.20\pm0.28$ \\
 { EW}(H$\beta$)$\dagger$ [\AA] & $112\pm13$ & $176\pm16$ \\
 \hline
\end{tabular}
 \label{tab:physical conditions}
 \begin{description}
     \item[] Note: O32 = [\ion{O}{iii}]/[\ion{O}{ii}] (a proxy for the the ionisation parameter).\\ 
     $^{\star}$ In these cases, the \TO\ and C/O were determined using the \Te\ UV-diagnostic of \ion{O}{iii}]~\W1666/[\ion{O}{iii}]~\W5007 and the \ion{C}{iii}]~\W1909/\ion{O}{iii}]~\W1666 line ratio respectively. Reassuringly, in both cases the values are fully consistent with value derived using the optical [\ion{O}{iii}]~\W4363 line. {$\dagger$ EW(H$\beta$) in the Rest-frame. }
 \end{description}
\end{table}

\subsection{Physical conditions and chemical abundances}
\subsubsection{Electron density and temperature}

The electron temperature (\Te) and density (\Ne) for EXCELS-70864 and EXCELS-121806 were determined using the nebular analysis package \textsc{PyNeb} (\citealp{luridiana15}; version 1.1.14) using the same atomic data reported in \citet{arellanocordova24}. 
All atomic data assume a five-level atom model \citep{derobertis87} with the exception of O that uses a six-level atom. 
For C, we use the transition probabilities of \citet{glass83}, \citet{nussbaumer78}, and \citet{wiese96} and the collision strengths of \citet{berrington85}. 

We have used the [\ion{O}{iii}]~\W4363/\W5007 and [\ion{S}{ii}] \W6717/\W6731 line ratios to calculate \Te[\ion{O}{iii}]\ and \Ne[\ion{S}{ii}], respectively. 
For EXCELS-121806, we have also calculated \Te[\ion{O}{iii}]\ using the UV auroral line [\ion{O}{iii}]~\W1666 via the [\ion{O}{iii}]~\W1666/\W5007 ratio. 
The electron density is only constrained for EXCELS-121806, for which we find a value of $n_e={600}^{+920}_{-400} \, \mathrm{cm}^{-3}$; we adopted the same value for EXCELS-70864.
In Table~\ref{tab:physical conditions}, we report the electron density and temperature for both galaxies. 
We find values of \TO $= 12500^{+\,1100}_{-1300} \, \rm{K}$ and \TO $= 14900\pm1100 \, \rm{K}$ for EXCELS-70868 and EXCELS-121806, respectively. 
Our constraint on \TO\ using the UV+optical line ratios is also in agreement with the result using only optical lines, \TO\ (UV) $= 15300\pm1250 \, \mathrm{K}$. 
These values are consistent with the inferred \Te\ in the CLASSY low-redshift analogue sample \citep[e.g.,][]{mingozzi22, arellanocordova24}. 
In this study, we used the result obtained for \TO\ using optical lines to derive the ionic abundances.

\subsubsection{Ionic and total abundances}\label{sub_sec:ionic}
We have defined a three ionisation zone temperature structure to determine the ionic chemical abundances. 
We use \TO\ as representative of the high ionisation zone, and \To\ (or \TN) and \TS\ as a representative of the low and intermediate ionisation zones, respectively. 
However, \To, \TN\ and \TS\ are not available in this sample. 
Therefore, we have used temperature relations to estimate the low- and intermediate- ionisation zone temperatures.
We have estimated \To\ using the temperature relation reported in \cite{arellanocordova20}, which is based on observational data and accounts for the dependency on the ionisation parameter. 
To estimate \TS, we have used the temperature relation of \citet{garnett92}, which is based on the photoionisation models of \cite{stasinska82}. 
Similar relations are used in many other studies of both local and high-redshift galaxies throughout the literature \citep[e.g.,][]{kennicutt03a,  arellanocordova20, Berg22, rogers24, arellanocordova24, hu24}. 

Based on the estimates of \Te\ and \Ne, we then derived the element abundances relative to hydrogen using the emission-line fluxes of the various ions observed in the EXCELS spectra. 
In addition to C, N and O, which are the main focus of this work, we also determined Ne and Ar abundances using the $[\ion{Ne}{iii}]$~\W3869 and $[\ion{Ar}{iii}]$~\W7135 lines where available (Table~\ref{tab:fluxes}) which we will discuss briefly below. 
We computed the ionic abundance of low-ionisation species (N$^+$ and O$^+$) using \To, intermediate-ionisation species (Ar$^{2+}$) using \TS, and high-ionisation species (C$^{2+}$, O$^{2+}$, and Ne${^{2+}}$)\footnote{{ \TS\ is also used to infer C$^{2+}$ due to the similarity in ionisation potentials between S$^{2+}$ (23-35 eV) and C$^{2+}$ (24-48 eV). We recalculated the C/O ratios using \TS\, finding differences smaller than 0.09 dex compared to the values reported in Table~$\ref{tab:physical conditions}$. These differences lie within the uncertainties associated with the C/O ratios derived using \TO\ \citep[see also][]{garnett92, jones23}.}} using \TO\ {\citep[e.g.,][]{garnett92, jones23}}. 
For O, we also derived the ionic abundance of O$^{2+}$ using \ion{O}{iii}]~\W1666 and [\ion{O}{iii}]~\W5007.

These ionic abundances were then converted to total element abundance using various prescriptions.
For O, the total abundance was calculated by summing the contribution of $\rm{O}^{+}/\rm{H}^{+}$ and $\rm{O}^{2+}/\rm{H}^{+}$. 
The contribution of $\rm{O}^{3+}$/H$^{+}$ is expected to be minimal even in highly-ionised environments and therefore we do not consider any additional ionisation correction factor \citep{berg21a}. 
We derive total gas-phase oxygen abundances (i.e., metallicities) of ${12+\mathrm{log(O/H})=8.21^{+0.07}_{-0.05}}$ for EXCELS-70864 and ${12+\mathrm{log(O/H})=7.96^{+0.05}_{-0.04}}$ for EXCELS-121806. 
These values translate into metallicities of $Z \simeq 0.3 \, \mathrm{Z_{\odot}}$ and $Z \simeq 0.2 \, \mathrm{Z_{\odot}}$ respectively.
As discussed below, these estimates are at the upper end of the derived values for other galaxies at $z\gtrsim5$ in the literature \citep[e.g.,][]{isobe23b, morishita24}. 

For all of the other elements, we used the ionisation correction factors (ICFs) to take into account the contribution of the unobserved ions. 
For N, we have used the ICF proposed by \cite{peimbert69}, which is based on the similarity of the ionisation potentials of $\mathrm{N}^{+}$ and $\mathrm{O}^{+}$, such that $\mathrm{N}^{+}/\mathrm{O}^{+} \approx \mathrm{N/O}$.
To derive the total C abundances, we use the ICF of \cite{Berg22}, which is valid for high-ionisation galaxies, and has been used in other studies at high-redshift \citep[e.g.,][]{arellanocordova22b, jones23,citro24}. 
Recently, \citet{izotov23} provided a bespoke ICF for C based on the properties of LCE at $z=0$; we also tested this ICF, and obtained very similar results with a difference of only 0.03 dex.
For Ne and Ar, we follow the recommendation of \cite{arellanocordova24} who analysed the ICFs for several ions using the CLASSY sample at $z\simeq0$ and recommend the ICFs of \cite{Dors13} for Ne, and \citet{izotov06} for Ar. 

To calculate the uncertainties associated with the ISM conditions and chemical abundances, we used Monte Carlo simulations generating a Gaussian distribution of 1000 random values. 
The final values are taken as the median of the resulting distribution, and the reported uncertainties are determined from the 16$^{\mathrm{th}}$ and 84$^{\mathrm{th}}$ percentiles. Additionally, we have added in quadrature an uncertainty of 610 K and 1300 K to the estimated of \TS\ and \To, respectively \citep[e.g.,][]{Rogers21, Stanton25}.
The abundances and corresponding uncertainties for our two $z\simeq5$ galaxies are reported in Table~\ref{tab:physical conditions}.

\subsubsection{Neon and argon abundances}

We also report the $\rm{Ne/O}$ and $\rm{Ar/O}$ abundance ratios in Table~\ref{tab:physical conditions}. 
Our derived values for $\rm{Ne/O}$ are in excellent agreement with the Solar ratio ${(\mathrm{log(Ne/O)}_{\odot} = -0.63)}$ and with the average value of ${\mathrm{log(Ne/O)} = -0.63\pm0.06}$ reported \citet{arellanocordova24} for the $z\simeq0$ CLASSY sample \citep[see also][]{izotov06, berg19a}.  
Furthermore, a solar-like $\rm{Ne/O}$ has also been reported for a number of other high-redshift samples \citep[e.g.][]{arellanocordova22b, marqueschaves23, schaerer24}.
Indeed, a solar-like $\rm{Ne/O}$ ratio is expected at high-redshifts as both are $\alpha$-elements, and their ratio at low metallicities should be determined by the CCSNe yields of massive stars \citep{kobayashi20}.  
The consistency of our data with the Solar value, and with other local and high-redshift systems, is therefore reassuring evidence that our flux calibration and dust corrections are likely to be reasonable.

For EXCELS-121806 we have a measurement of the [\ion{Ar}{iii}] \W7135 line from which we derived a value of ${\mathrm{log(Ar/O)} = -2.20\pm0.28}$ using the ICF of \citet{izotov06}.
This value is slightly higher than observed in local SFGs \citep[e.g.,][]{izotov06, arellanocordova24}; for example, the mean value reported for the CLASSY galaxies is ${\mathrm{log(Ar/O)} =-2.33\pm0.09}$ \citep{arellanocordova24}, which is consistent with the Solar value of \cite{asplund21}. 
To check the dependence on the ICF, we also used the ICF of \cite{perezmontero07} which yields a lower value ${\mathrm{log(Ar/O)} = -2.40\pm0.28}$, but is still fully consistent within the uncertainties.
In a companion work, \citet{Stanton25} determined the average $\rm{Ar/O}$ abundance ratio for nine EXCELS galaxies $z=3-5$ (which includes EXCELS-121806) finding a weighted average value of $\langle {\mathrm{log(Ar/O)} \rangle =-2.52\pm0.07}$, consistent with an Ar deficit due to the delay in  Type Ia supernovae enrichment at high-redshifts.
Given the large uncertainty, it is difficult to draw any firm conclusions from this single $z=5$ $\mathrm{Ar/O}$ abundance measurement, but the value we derive is clearly within the range of expected values. Finally, we revised the dependence of abundance determinations on electron density by varying \Ne = 10$^{2}$-10$^{4}$ cm$^{-3}$. High electron densities have been reported both locally and at high-$z$ using various UV diagnostics that trace different desnity gas structure within the nebula \citep[e.g.,][]{mingozzi22, topping24a}. Overall, we find that variations in O/H, C/O, Ne/O, and Ar/O remain below 0.1 dex compared to the values reported in  Table~\ref{tab:physical conditions}. For N/O, we note a large discrepancy in such a ratio with a difference of 0.33 dex by assuming \Ne = 10$^{4}$ cm$^{-3}$. Such high densities cannot be confirmed as we lack detections of other diagnostics that trace high-density diagnostics such as [\ion{Ar}{IV}]~\W4711,41, while that \ion{C}{iii}]~\W\W1906,09 is blended (see Fig.~\ref{fig:spec_1}).

\subsubsection{A comparison to strong-line methods}

It is perhaps useful to consider what values of $\mathrm{O/H}$, $\mathrm{N/O}$ and $\mathrm{C/O}$ we would obtain if we did not have an estimate of the electron temperature and had to instead rely on empirical strong line methods.
To do this, we first inferred the metallicity of EXCELS-70864 and EXCELS-121806 using the strong-line calibrations of \citet{sanders24}, which were derived using a sample of galaxies at $z=2-9$ with \Te-measurements.
We used the emission-line ratios O2 ($=$[\ion{O}{ii}]/H$\beta$), O3 ($=\,$[\ion{O}{ii}]/H$\beta$), R23 ($=\,$([\ion{O}{ii}] + [\ion{O}{iii}])/H$\beta$), O32 ($=\,$[\ion{O}{iii}]/[\ion{O}{ii}]), and Ne3O2 ($=\,$[\ion{Ne}{iii}]/[\ion{O}{iii}]). 
We did not consider calibrations using the [\ion{N}{ii}] line due to their secondary dependence on the N/O ratio \cite[e.g.,][]{arellanocordova20}. 
We averaged the metallicities derived from the different calibrators to compare with the results of the direct method.
For EXCELS-121806, we find $\rm 12+log(O/H) = 7.92$, which is 0.05 dex lower than the value obtained using the \Te-sensitive method but fully consistent within the uncertainties (see Table~\ref{tab:physical conditions}). 
For EXCELS-70864, we infer $\rm 12+log(O/H) = 7.87$, which is 0.34 dex lower than our \Te-based estimate.
We have also checked that applying the reddening correction of $E(B-V) = 0.19$ derived from the higher-order Balmer lines for EXCELS-70864 results in a similar offset.
This offset is larger than the scatter on the O2, O3, Ne3O2, O32, and R23 methods \citep{sanders24} and highlights the fact that strong line methods are likely to artificially reduce the true scatter in abundance measurements at high-redshifts and are not reliable when applied to individual objects.

\begin{figure*}
\begin{center}
    \includegraphics[width=0.98\textwidth, trim=5mm 0mm 0mm 0mm, clip=yes]{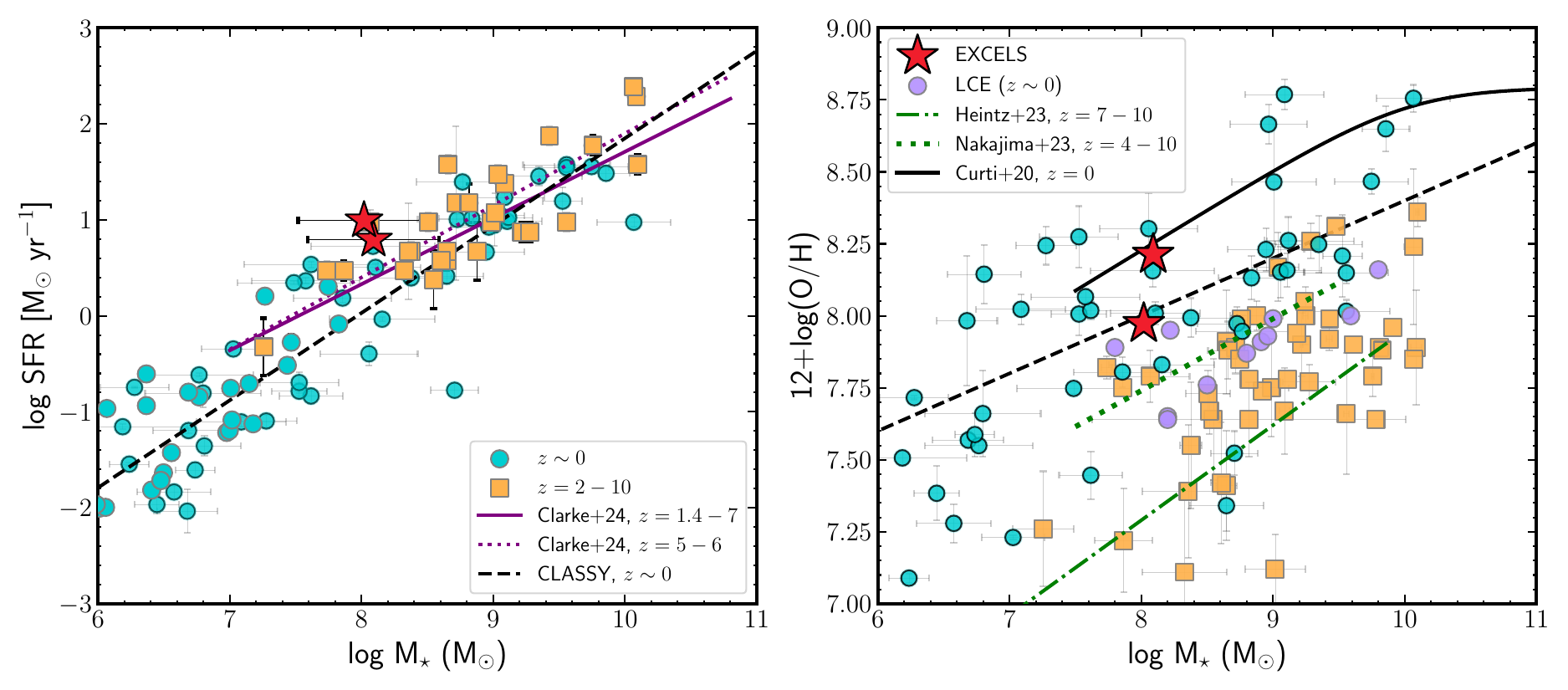}
        \caption{{\it Left:} The SFR versus stellar mass relation for the EXCELS galaxies at $z\simeq5$ (red stars), along with an additional sample from the literature. Star-forming galaxies at $z\simeq0$ galaxies (circles) are taken from \citet{berg16, berg19a, izotov23, Ravindranath20, senchyna21, Berg22}, and at $z>2$ (squares) from \citet[][]{sanders23b, morishita24, marqueschaves23, citro24, jones23, arellanocordova24, llerena23,schaerer24, isobe23a, rogers24, topping24a, topping24b, curti24b}. 
        The solid and dotted lines shows the SFR-M$_{\star}$ relation from \citet{clarke24} at $1.4 < z < 7$ and $5<z<6$, respectively. 
        The dashed line shows the SFR-M$_{\star}$ relation for the CLASSY sample \citep{Berg22}. 
        The high redshift galaxies and local analogues fall on the same SFR-M$_{\star}$ sequence despite the clear differences in metallicity. 
        {\it{Right}}: The mass-metallicity relationship (MZR) for the same sample including LCE galaxies distinguished from other local SFGs for comparison purposes \citep[][purple circles]{izotov23}. 
        The MZR for the CLASSY high-redshift analogue sample is shown as the dashed line \citep{Berg22} while the solid line represents the average MZR for all local SFGs \citet{Curti20}. 
        In addition, two MZRs from high-redshift studies are indicated with dotted and dot-dashed lines from \citet[][]{nakajima23} ($z =4-10$) and \citet[][]{heintz23} ($z=4-10$), respectively. We find that the two EXCELS galaxies $z\simeq5$ are metal-rich in comparison to other galaxies at similar redshifts, and are in fact consistent with the CLASSY MZR.} 
\label{fig: SFR_comparison}
\end{center}
\end{figure*}

\subsection{Stellar masses and star-formation rates}
\label{sec:galaxy_properties}

To estimate stellar masses and star-formation rates, we fitted the broadband photometry of the two EXCELS galaxies using the {\sc{BAGPIPES}} SED fitting code \citep{carnall18, carnall19a}.
In addition to the \emph{HST} and \emph{JWST}/NIRCam photometry described above, EXCELS-121806 falls within the PRIMER/MIRI region and therefore benefits from additional \emph{JWST}/MIRI $7.7 \, \mu \mathrm{m}$ photometry which we also include in the SED fitting for this object.
We first subtracted the measured emission-line fluxes from the photometric data to ensure that we were fitting only to the stellar and nebular continuum SED components.
We used the \citet{bruzual03} models, and assumed a flexible dust attenuation law following \citet{salim2018} and a double power-law star-formation history with a \citet{kroupa01} IMF; nebular continuum emission was included in the fitting but emission lines were excluded.
Full details of the \textsc{Bagpipes} fitting for the EXCELS star-forming sample are given in \citet{Scholte25}.
The resulting stellar masses and SFRs are listed in Table~\ref{tab:physical conditions}.
We derive stellar masses of ${\mathrm{log}(M_{\star}/\mathrm{M}_{\odot}) = {8.09}^{+\, 0.24}_{-0.15}}$ and ${\mathrm{log}(M_{\star}/\mathrm{M}_{\odot}) = { 8.02}^{+\, 0.06}_{-0.08}}$ for EXCELS-70864 and EXCELS-121806, respectively. 
These stellar masses are similar those of other $z>6$ galaxies (e.g., \citealp{curti24b, Hsiao24b};  see also Sec~\ref{sec:MZR-SFR}).
In addition to returning the best-fitting physical properties, these SED fits also returned an estimate of the stellar continuum SED which we used to correct the Balmer line fluxes for underlying stellar absorption (see \citealp{Scholte25}).
The Balmer absorption corrections were small for both EXCELS-70864 and EXCELS-121806 ($< 2\%$ ).

We also derived the SFR for both galaxies using the observed 
H$\alpha$ fluxes (see Table~\ref{tab:fluxes}). 
The luminosity distances for EXCELS-70864 and EXCELS-121806 were calculated based on the redshifts reported in Table~\ref{tab:physical conditions} and used to convert the $\mathrm{H}\alpha$ fluxes to luminosities, $L(\mathrm{H}\alpha)$.
The SFR was estimated using the relation between SFR and $L(\mathrm{H}\alpha)$ reported in \citet{reddy18} with a conversion factor of 3.236$\times10^{-42}$ {  erg s$^{-1}$}, which is appropriated for low-metallicity systems. 
 We derived values of $\mathrm{log(SFR/M_{\odot}\mathrm{yr}^{-1})} = 0.79\pm0.01$ for EXCELS-70864 and $\mathrm{log(SFR/M_{\odot}\mathrm{yr}^{-1})} = 1.00\pm0.01$ for EXCELS-121806. 
In both cases, we find excellent agreement between the SFR derived from the SED fitting and the H$\alpha$ flux.

In the left-hand panel of Fig.~\ref{fig: SFR_comparison} we show the position of the EXCELS galaxies in the star-forming main sequence plane compared to the sample from the literature. 
The different lines correspond to the relations between SFR and stellar mass for galaxies at $z\sim0$ from the CLASSY survey (\citealp{Berg22}; dashed line) and galaxies at a similar redshift (\citealp{clarke24}; dotted and solid lines). 
It can be seen that the two EXCELS galaxies agree well with the relation derived for the $z\sim0$ CLASSY sample, the additional sample of galaxies at $z\geq2$, and the relations from \citet{clarke24}.
Therefore, the global properties of the local and high-redshift samples are clearly comparable.


\section{Chemical abundance patterns}
\label{sec:CNO_results}

In this section, we first present the total metallicity (i.e., oxygen abundance) of our two $z\simeq5$ galaxies before investigating the N/O and C/O ratios as a function of metallicity and stellar mass.
Throughout we compare with our local and high-redshift comparison samples.
Finally, we compare our results with bespoke chemical evolution models following the \citet{kobayashi20} prescription to explore the potential enrichment scenarios for our two galaxies. 

\begin{figure*}
\begin{center}
    \includegraphics[width=0.95\textwidth, trim=5mm 0mm 0mm 0mm, clip=yes]{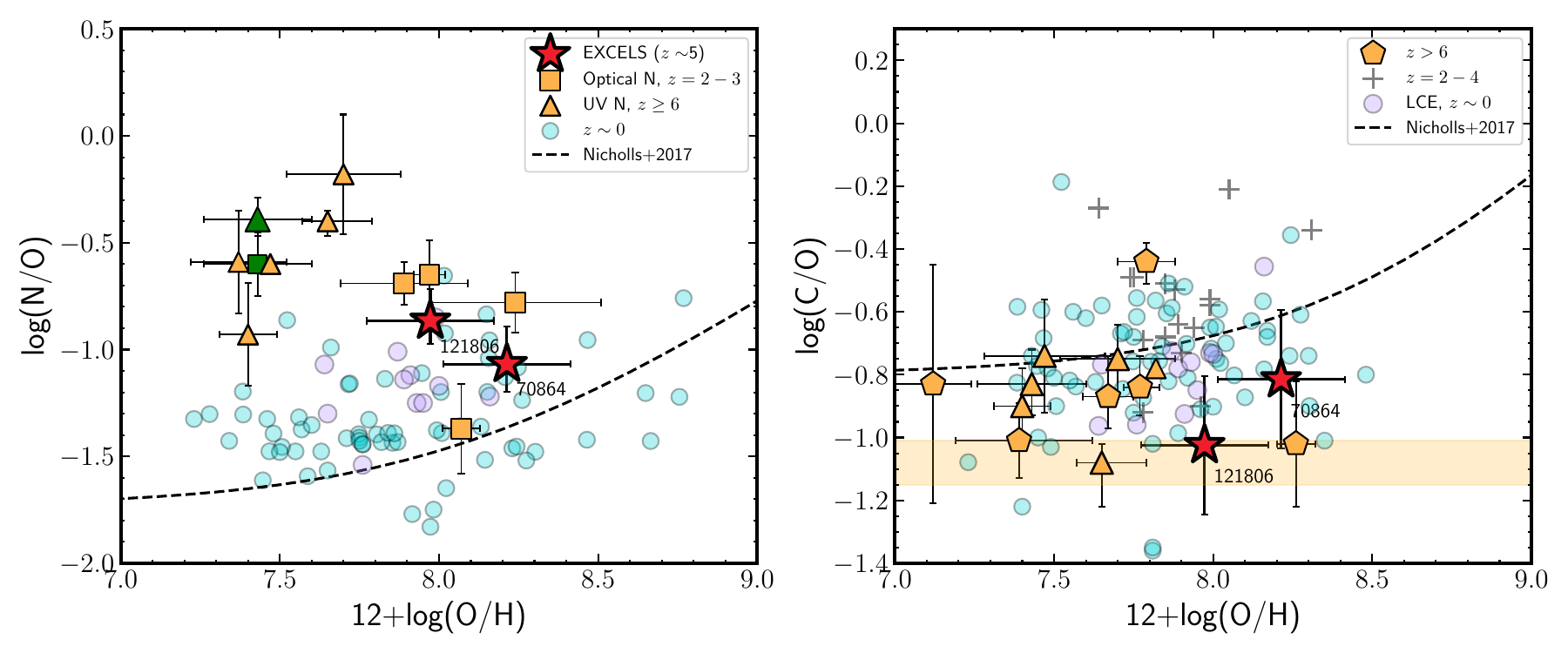}
        \caption{{\it Left-hand panel:} The $\rm{N/O}-\rm{O/H}$ values for the EXCELS galaxies (red stars) compared with SFGs at different cosmic epochs. 
        The circles represent galaxies at $z\simeq0$ \citep[][Arellano-Cordova et al. in preparation]{berg16, berg19a, Ravindranath20, senchyna21, Berg22, izotov23}, while squares shows galaxies at $z=2-3$ with N/O determined using optical lines from \citet[][]{sanders23a} ($z\sim2$), \citet[][]{rogers24} (Q2343-D40; $z\sim2.96$) and \citet[][]{welch24b} (Sunburst Arc; $z=2.37$). 
        SFGs with N/O determined using UV emission-lines (\ion{N}{iii}] and/or [\ion{N}{iv}]) are shown as triangles. 
        LCE at $z\simeq0$ from \citet{izotov23} are depicted by purple circles. 
        Additionally, our N/O result using only optical lines from \citet{topping24a} for RXCJ2248-ID is included as a green triangle, while the UV N/O is shown as a green square.
        The EXCELS galaxies fall above the mean value of $\rm{N/O}$ for the local CLASSY sample and there appears to be a general trend whereby galaxies at $z>2$ lie systematically above the local relation.
        {\it Right-hand panel:} A similar comparison in the $\rm{C/O}-\rm{O/H}$ plane. 
        Galaxies with both C and N abundances (i.e. that appear in both panels) are represented by triangles. 
        We also include C/O measurements from \citet{llerena23} at at $z=2-4$. 
        The yellow shading represents the C/O ratio from pure CCSNe enrichment derived using chemical evolution models in \citet{jones23}. 
        The EXCELS galaxies have higher $\rm{O/H}$ compared the with other high-redshift galaxies but show consistent $\rm{C/O}$ ratios.
        In both panels the dashed line shows the relations reported in \citet{nicholls17} for N/O and C/O based on Milky Way stellar abundances.}
\label{fig:CNO_excels}
\end{center}
\end{figure*}

\subsection{The gas-phase oxygen abundance}
\label{sec:MZR-SFR}

In the left-hand panel of Fig.~\ref{fig: SFR_comparison} we show the position of our two EXCELS galaxies in the mass-metallicity plane compared to our additional samples.
The metallicities we derive are 12+log(O/H) = $8.21^{+0.08}_{-0.05}$ for EXCELS-70864 and 12+log(O/H) = $7.97^{+0.05}_{-0.04}$ EXCELS-121806 (i.e., $Z \simeq 0.3 \, \mathrm{Z}_{\odot}$ and  $Z \simeq 0.2 \, \mathrm{Z}_{\odot}$, respectively.)
As a result, both galaxies are in fact consistent with the mass-metallicity relation (MZR) derived for the local CLASSY sample at $z\simeq0$ \citep{Berg22}.
At fixed stellar mass, they both have higher metallicities than the majority of the $z\geq2$ SFGs in our comparison sample, and the local LCE from \citet{izotov23}.
Remarkably, the metallicities are not far below the \emph{average} $z=0$ MZR from \citet{Curti20}.

In Fig.~\ref{fig: SFR_comparison}, we have also added the MZRs derived by \citet{heintz23} and \citet{nakajima23} for galaxies at $z = 7-10$ and $z = 4-10$ respectively, although we note that these MZRs are derived using strong-line calibrations to infer the metallicity. 
That being said, \cite{morishita24} have reported direct metallicity determinations for galaxies at $z\simeq3-9$ which are consistent with the relation of \citet{heintz23}.
A comparison to these high-redshift relations makes it clear that EXCELS-70864 and EXCELS-121806 are both metal rich for their stellar mass at $z\simeq5$, with metallicities that are directly comparable to local analogues from the CLASSY sample; these results again highlight the potentially large scatter in the metallicities of high-redshift galaxies that is only revealed using direct $T_e$-based metallicity constraints.

These relatively high metallicities offer the opportunity to place new unique constraints on CNO abundance patterns in high-redshift galaxies.
As we discuss below, most galaxies at $z\gtrsim5$ with direct C, N and O abundance estimates have $\mathrm{12+log(O/H)} \lesssim 7.8$; our observations will therefore place the first constraints on comparatively `metal-rich' galaxies in the early Universe.

\subsection{The $\mathrm{\mathbf{N/O}}$ abundance ratio}

In the left-hand panel of Fig.~\ref{fig:CNO_excels}, we show the N/O abundance ratio of our two galaxies as a function of metallicity. 
It can be seen that, compared to the $z\simeq0$ samples, both $z\simeq5$ galaxies fall within the upper envelope of $\mathrm{N/O}$ at fixed $\mathrm{O/H}$\footnote{We also infer the N/O ratio using strong-line methods from \cite{perez-montero09}, \citet{hayden-pawson22}, and \citet{florido22}. These methods are mainly based on samples of local SFGs and \ion{H}{ii} regions. For EXCELS-70864 and EXCELS-121806, we infer average values of log(N/O) = $\mathrm{-1.13}$ and log(N/O) = $\mathrm{-1.02}$, respectively. These values are 0.06 dex and 0.16 dex lower than those calculated using the \Te-sensitive method, but they are in agreement within the uncertainties.}.
Therefore, in contrast to some of the highly N-enriched galaxies that have been reported at $z > 6$ \citep[e.g.,][]{Bunker23, marqueschaves23}, the N enrichment of our sources is not unprecedented compared to what is observed locally.
This difference might be explained by the fact that the metallicity of our sample is slightly larger than the typical metallicity of these notable N-rich galaxies (see Fig.~\ref{fig:CNO_excels}).
In fact, when compared to star-forming galaxies at slightly lower redshift ($z =2-3$; orange squares) which have similar $\mathrm{O/H}$, it can be seen that our $\mathrm{N/O}$ measurements are generally consistent within the uncertainties \citep[][]{sanders22, welch24b, rogers24, citro24}. 
 
However, despite the fact that some $z\simeq0$ galaxies have similar $\mathrm{N/O}$ at fixed $\mathrm{O/H}$, there does appear to be a general trend for the high-redshift galaxies to be more N-enriched than the typical local sources (Fig.~\ref{fig:CNO_excels}).
Of our two $z\simeq5$ galaxies, EXCELS-121806 in particular is clearly more N-rich than the majority of the $z=0$ sample.
In general, if we consider all of the $z>2$ sources, $12/14$ ($85$ per cent) have $\mathrm{log(N/O) > -1}$ compared to only $8/58$ ($14$ per cent) of the local samples.
Of these local sources, it is clear that the LCEs of \citet{izotov23} have systematically higher $\mathrm{N/O}$ compared to the CLASSY sample, which \citet{izotov23} conclude is an indication of additional sources of N enrichment in these highly ionised galaxies. 
Still, the majority of $z>2$ galaxies, including our two $z\simeq5$ EXCELS sources, display a higher $\mathrm{N/O}$ than the typical $z=0$ LCE.

If we compare directly to the $z\simeq2-3$ galaxies with $\mathrm{12+log(O/H)} \simeq 8.0$, to which our two EXCELS galaxies seem most comparable, there is evidence for a potentially large scatter in $\mathrm{N/O}$ in high-redshift galaxies. 
\citet{sanders23a} reported an elevated N/O ratio (using [\ion{N}{ii}] \W6584) for two high specific SFR (sSFR) galaxies at $z\simeq2$ ${(\mathrm{sSFR} \simeq 20 \, \mathrm{Gyr}^{-1})}$ and attribute the $\mathrm{N/O}$ enhancement to the accretion of pristine gas which dilutes the overall metallicity content while preserving $\mathrm{N/O}$. 
Similarly, \citet{welch24b} recently reported the abundance patterns of different elements derived from a stacked spectrum of the Sunburst Arc galaxy at $z$= 2.37, finding a N/O enrichment for this galaxy which they attribute to the possible signature enrichment of WR stars.
On the other hand, \citet{rogers24} report the N/O ratio of a $z\simeq3$ galaxy (Q2343-D40) that is in good agreement with the average $\mathrm{N/O-O/H}$ relation observed locally (see Fig.~\ref{fig:CNO_excels}). 

Overall, the comparison in the left panel of Fig.~\ref{fig:CNO_excels} suggests that the $\mathrm{N/O}$ ratios at fixed $\mathrm{O/H}$ of our $z\simeq5$ galaxies are not unprecedented when compared to the local sample.
Despite this, they do seem to follow a general trend in which galaxies at ${z \gtrsim 2-3}$ fall systematically above the average local $\mathrm{N/O}-\mathrm{O/H}$ relation.
This systematic offset suggests that the high-redshift population seems to be, on average, more N-rich than their local counterparts.
More samples of direct-method N measurements at $z \gtrsim 5$ are needed to robustly demonstrate this claim.

\subsection{Inferring $\mathrm{\mathbf{N/O}}$ from UV emission-lines}

One potentially important difference between estimates of $\mathrm{N/O}$ above and below $z\simeq6$ is the fact that, at $z>6$, the abundances of N are determined primarily using UV emission-lines as opposed to the optical [\ion{N}{ii}]~\W6584 line we have used.
Indeed, the unexpected detection of the \ion{N}{iii}] \W\W1747-1750 (quintuplet) or \ion{N}{iv}\W\W1483,1486 lines with  JWST has opened a new pathways to the analysis of the different ionisation states of N and their ionic and total abundances at $z>6$ \citep[e.g.,][]{Bunker23, Cameron23}. 
Several studies have reported significant N/O enrichment in galaxies at $z>6$ based on these lines (\citealp{Bunker23, Cameron23, senchyna23, marqueschaves23, topping24a, topping24b, schaerer24, curti24b}; see Fig.~\ref{fig:CNO_excels}).
These results imply an additional enrichment to understand the high N/O in the early Universe \citep[e.g.,][]{charbonnel23, senchyna23, kobayashi24}. 

\citet{topping24a} presents an analysis of C and N in a lensed galaxy at $z = 6.1$ (RXCJ2248-ID). 
Interestingly, the spectrum of RXCJ2248-ID provides information on three different ionisation states of N: [\ion{N}{ii}] in the optical and \ion{N}{iii}] and [\ion{N}{iv}] in the UV. 
\cite{topping24a} derived the N/O abundance ratio using only the UV lines, resulting in a very high value of N/O (see Table~\ref{tab:high-z_sample}). 
Since in \cite{topping24a} they do not report the $\mathrm{N/O}$ ratio using the optical [\ion{N}{ii}] line, we have used the [\ion{N}{ii}]~\W6584 flux listed in \cite{topping24a} to derive the abundance ratio of N/O following the same method applied to our $z\simeq 5$ sources (we report this value in Table~\ref{tab:high-z_sample}). 
In Fig.~\ref{fig:CNO_excels}, we show the two values of $\mathrm{N/O}$ derived using either the optical N lines (green square) or the UV N lines (green triangle) for RXCJ2248-ID. 
The optical-based $\mathrm{N/O}$ ratio (green square) is 0.21 dex lower than UV-based estimate, although consistent within the uncertainties. {In addition, by taking advantage of the three different ionisation states of N for RXCJ2248-ID, we also infer the N/O ratio without using an ICF assuming that N/O = (N$^{+}$ + N$^{2+}$ + N$^{3+}$)/(O$^{+}$ + O$^{2+}$). Therefore, we have calculated a value of log(N/O) $= -0.52^{-0.18}_{+0.06}$, which is 0.08 dex higher than optical N/O reported in Table~\ref{tab:high-z_sample}. This value is also consistent with the N/O derived using log(C$^{++}$/N$^{++}$ in combination with log(C/O) present in \cite{topping24a} \citep[see also][]{berg18, curti24b}. 
Despite the slightly lower value, the optical $\mathrm{N/O}$ estimate remains extremely high at fixed $\mathrm{O/H}$ compared to local galaxies.

This comparison suggests that there is no significant bias in the $\mathrm{N/O}$ ratios in the optical and UV; the lower $\mathrm{N/O}$ of our sample compared to the $z>6$ samples in Fig.~\ref{fig:CNO_excels} is likely not a result of the differences in the lines used to calculate the N abundance. 
However, this is only one example, and it should be kept in mind that the N/O ratio derived from UV lines might be sensitive to different ISM conditions (i.e,, high-density gas; \citealp{pascale23, senchyna23}) that could, in principle, systematically bias UV and optical estimates. 
Larger samples of combined UV and optical N abundance measurements are needed to fully explore this issue.

\begin{figure*}
\begin{center}
    \includegraphics[width=0.8\textwidth, trim=5mm 0mm 0mm 0mm, clip=yes]{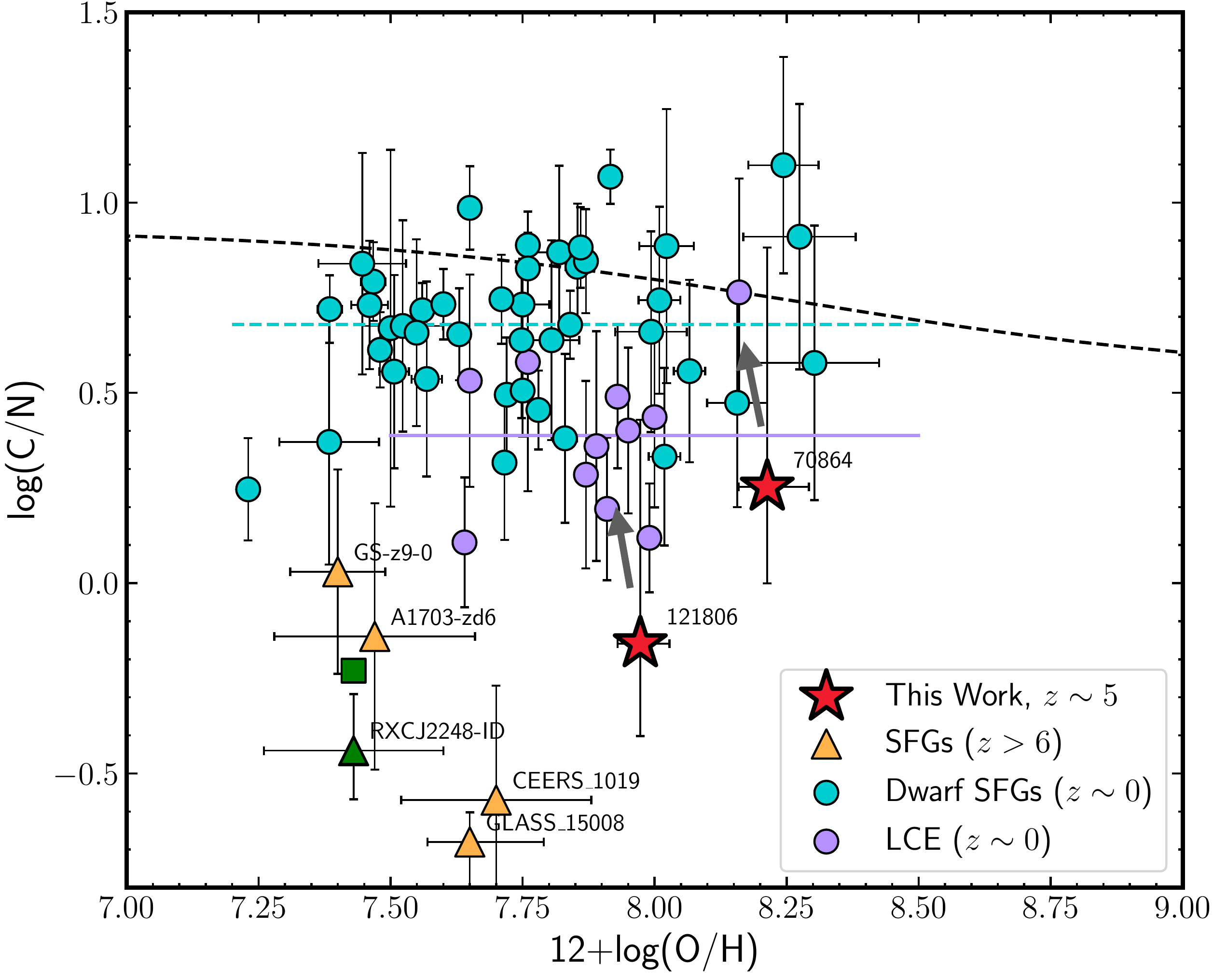}
        \caption{The $\rm C/N-O/H $ relationship for the EXCELS galaxies at $z\simeq5$ compared to literature samples. 
        The sample of local dwarf SFGs and LCEs are represented with circles (\citealp{berg12, berg16, izotov23}; Arellano-C\'ordova et al. in prep; Martinez et al. in prep).
        The blue dashed and purple solid lines show the mean of two local dwarf SFGs and LCEs, respectively. 
        The triangles show the high-redshift galaxies from \citet[][]{isobe23b} (GLASS$\_$15008; $z=6.23$), \citet[][]{marqueschaves23} (CEERS$\_$1019; $z=8.63$), \citet[][]{topping24a} (RXCJ2248-ID; $z=6.11$), \citet[][]{topping24b} (A1703-zd6; $z=7.04$), \citet[][]{schaerer24} (GN-z9p4, $z=9.38$), \citet[][]{curti24b} (GS-z9-0, $z=9.43$). 
        The $\rm{C/N}$ ratio that we have derived using the optical [\ion{N}{ii}] \W6584 line reported in \citet[][RXCJ2248-ID]{topping24a} is shown as the green square (see Section~\ref{sec:CNO_results}). 
        The EXCELS galaxies and other high redshift galaxies fall systematically below the average $\rm{C/N}$ ratios of the local analogues. 
        The grey arrows indicate dust vector ({the position of the EXCELS galaxies}) for assuming small amounts of reddening indicated by the higher-order Balmer line ratios (see Sec.~\ref{sec:methodology}). 
        The dashed line represents the relation from \citet{nicholls17} based on Milky Way stellar abundances.} 
\label{fig:CN_relation}
\end{center}
\end{figure*}

\subsection{The $\mathrm{\mathbf{C/O}}$ abundance ratio}

We now turn to the C abundances of our two $z\simeq5$ galaxies.
We have calculated ${\mathrm{log(C/O)} = -0.82\pm0.22}$ and ${\mathrm{log(C/O)} = -1.02\pm0.22}$ for EXCELS-70864 and EXCELS-121806 respectively.  Using only the UV lines we reassuringly find a fully consistent value of ${\mathrm{log(C/O)} = -1.07\pm0.43}$ (see Table~\ref{tab:physical conditions}).
In the right-hand panel of Fig.~\ref{fig:CNO_excels} we show the $\mathrm{C/O}$ versus $\mathrm{O/H}$  abundance ratios for the EXCELS galaxies compared to the additional literature samples. 
As in the left-hand panel, the local LCE are identified as purple circles to distinguish them from the local dwarf SFGs from CLASSY. 
Galaxies which have both $\mathrm{C/O}$ and $\mathrm{N/O}$ abundance estimates are shown as triangles; galaxies with only C/O measurements are identified with pentagons. 
It can be seen that, in contrast to $\mathrm{N/O}$, the $\mathrm{C/O}$ ratios of the EXCELS galaxies - and the other high-redshift sources - fall systematically below the local SFGs.

Our two new measurements suggest these low $\mathrm{C/O}$ ratios extend to higher metallicities in agreement with the result of \citet{citro24}, who analyzed a lensed galaxy at $z= 3.77$ (J0332-3557) with a similar metallicity and find a similarly low $\mathrm{C/O}$ abundance. \citet{citro24} concluded that the low C/O is probably due to the enrichment of massive stars and low star-formation efficiency.

These low $\mathrm{C/O}$ abundance ratios imply that the source of enrichment is likely a result of the rapid enrichment of massive stars or other exotic scenarios, for instance, very massive stars \citep[e.g.,][]{watanabe24}.
This scenario would be consistent with the young ages derived from SED modeling ($< 100$ Myr) which would mean that enrichment of C via low-mass stars ($1-4 \, \mathrm{M}_{\odot}$; \citealp{henry00, kobayashi20}) has not had time to take place. 
The additional sample of galaxies at $z > 6$ shows similarly low $\mathrm{C/O}$ for metallicities between ${\mathrm{12+log(O/H)}=7.12-7.70}$ \citep[e.g.,][]{arellanocordova22b, jones23, topping24a, isobe23b,marqueschaves23}. 
At these low metallicities, there is better overlap between the high-redshift and local samples (see Fig.~\ref{fig:CNO_excels}), suggesting a common source of C enrichment at the same time as a significant differences in N enrichment in these sources.
At the higher metallicity of our sample, systematic offsets in both $\mathrm{N/O}$ and $\mathrm{C/O}$ seem to be present, although large sample sizes are clearly needed at ${\mathrm{12+log(O/H)} > 8}$.

Focusing briefly on other literature analyses, \citet{jones23} measured $\mathrm{log(C/O)} = -1.01$ for GLASS$-$15008 ($z=6.23$) and suggest that the C enrichment can be explained purely by massive-star CCSNe. 
Similarly, \cite{curti24b} analysed the $\mathrm{C/O}$ enrichment of a galaxy at $z = 9.38$ using chemical evolution models, and again concluded that the $\mathrm{C/O}$ enrichment is consistent with the yields from massive-star CCSNe.
On the other hand, \citet{Hsiao24b} recently reported a high ${\mathrm{log(C/O) = -0.44^{+0.06}_{-0.07}}}$ for a galaxy at $z=10.17$ (MACS0647-ID) at relatively low metallicity (see Table~\ref{tab:high-z_sample}). These authors concluded that it might be associated with the enrichment of carbon from intermediate-mass stars that start to appear, given the age ($>$ 100 Myr) implied in one of the clusters of MACS0647-ID.   

Overall, it seems that the enrichment of $\mathrm{C/O}$ in our two EXCELS galaxies is most likely attributed to the yields of massive stars via CCSNe, before significant enrichment from low-mass AGB stars has begun to affect the ISM.
The same explanation can also be used to explain the low $\mathrm{C/O}$ ratios observed across the majority of high-redshift sources (barring the \citealp{Hsiao24b} measurement).
Nevertheless, additional samples of galaxies at $z\simeq5$ with \Te\ measurements are clearly still required to fully assesses O/H variations and improve our understanding of C production at early times. 

\subsection{The cosmic evolution of $\mathrm{\mathbf{C/N}}$}

Bringing the abundances of C, N and O together offers further insights into the star-formation timescales and enrichment pathways in star-forming galaxies \citep[e.g.,][]{henry00, berg19a}. 
For example, at low-metallicities (i.e., after the onset of star-formation), a plateau is expected based on the C and N yields of the most massive stars; this should be subsequently followed by deviations due to N-enrichment from intermediate mass stars, then C-enrichment from low-mass stars \citep[e.g.,][]{pena-guerrero17, berg19a, kobayashi20, izotov23}.

In Fig.~\ref{fig:CN_relation}, we show the $\mathrm{C/N-O/H}$ relation for our two EXCELS galaxies compared to the additional sample (where measurements of both C and N are available). 
In this element abundance ratio diagram, the different enrichment pathways of the high-redshift and local samples are clearly evident.
Broadly, there appears to be a separation between the local and high-redshift samples at ${\mathrm{log(C/N)} \simeq 0}$.
We find that $100$ per cent of the local sample fall above this value, compared to only $1/8$ (EXCELS-70864; $\simeq 10$ per cent) of the high-redshift sample.
As was evident from Fig.~\ref{fig:CNO_excels}, the high-redshift galaxies appear to be preferentially enriched in N compared to C; our two new measurements hint that this trend may extend to more metal-rich systems than have previously been observed.

In general, as can also be seen from Fig.~\ref{fig:CNO_excels}, we find that our $z\simeq5$ galaxies display $\mathrm{C/N}$ ratios that are closer to the local LCE than to the local CLASSY high-redshift analogue sample. 
We find values of $\langle {\rm log(C/N)}\rangle = 0.69\pm0.21$ for the local high-redshift-analogues (consistent with the mean value derived in \citealp{berg19a}) and $\langle {\rm log(C/N)}\rangle  = 0.38\pm0.20$ for the local LCE (both indicated by horizontal lines in Fig.~\ref{fig:CN_relation}).
Both of our $z\simeq5$ measurements fall below these average values, although much closer to the LCE.
Although it is not the focus of this work, we note that this kind of abundance ratio comparison between local LCE and high-redshift galaxies might offer an alternative pathway to understanding the escape fraction of ionising photons during the reionisation epoch.

At this point, it is worth highlighting the most important systematic uncertainty in our results, namely the effect of dust reddening.
In Fig.~\ref{fig:CN_relation} we have indicated with arrows the shift in the C/N and O/H abundances when adding relatively modest amounts of reddening ($E(B-V) = 0.15$).  We have used the extinction law of \citet[][$R_{V}=3.1$]{cardelli89}\footnote{As an additional inspection, we explore the impact of adopting the Small Magellanic Clouds (SMC) extinction law by \citet{Gordon03} on our results of the C/N ratios (see Fig.~\ref{fig:CN_relation}). We find that the resulting C/N ratios differ by 0.10-0.12 dex, which represents a shift of comparable magnitude to that obtained when using the extinction law of \citet{{cardelli89}}.} for dust correction.
It can be seen that the shift in $\mathrm{C/N}$ is potentially significant, while O/H remains similar; this large $\mathrm{C/N}$ shift is a result of the large wavelength separation between the C and N emission-lines ($\Delta \lambda_{\rm rest} \simeq  4500${\AA}). 
Crucially, any dust reddening moves the high-redshift samples closer to the local samples. 
Despite the relatively high metallicity of our $z\simeq5$ galaxies, we consider our estimate of $E(B-V) = 0$ to be robust; moreover, we note that some of the CLASSY galaxies with similar metallicity also display zero/negligible reddening.
Nevertheless, it is important to emphasize that deep spectra with accurate nebular reddening corrections are crucial for this kind of CNO abundance analysis.

\subsubsection{Chemical evolution models}

In order to gain a deeper understanding of the possible chemical enrichment pathways of these two EXCELS galaxies, we have constructed illustrative chemical evolution models following the prescriptions of \citet{kobayashi20} and \citet{kobayashi24} including WR stars. 
Our aim is to use these models to help us understand the $\rm{O/H}$ versus $\rm{C/N}$ relation shown in Fig.~\ref{fig:CN_relation}, but we emphasize that we are not attempting to find an exact match to the star-formation and chemical abundance history (doing so would be infeasible given the large uncertainties).

We consider a standard IMF with a slope of $x= 1.3$ in the mass range ${0.01 \, \rm{M}_{\odot} \leq M \leq 120 \, \rm{M}_{\odot}}$ \citep[Eq. 2]{kroupa01} . 
In addition, we investigate the use of a shallow (i.e., more top-heavy) and steep (i.e., more bottom-heavy) IMF by considering slopes of $\alpha-1=x=1.1$, $x=1.9$ and $x=2.3$. 
We consider a toy star-formation history that we consider plausible for $z=5$ galaxies and that yields the correct oxygen abundance at the redshift of our sources.
The star-formation history consists of an early rising phase followed by a strong burst $\simeq 1 \, \rm{Gyr}$ after the onset of star-formation in which the star-formation rate increases by $\simeq 2$ orders of magnitude. 
The assumed star-formation history for each of the IMFs is illustrated in Fig.~\ref{fig:models}.
For a shallower IMF, a larger mass-loss from massive stars makes the star-formation history slightly extended (the ejected material provides additional fuel for star formation).
As we use the same star-formation and infall timescales for these models ($3$ Gyr for the initial star formation, $0.1$ Gyr for the second star burst), the increase of stellar feedback for a shallower, more top-heavy, IMF is not accounted for here.

\begin{figure}
\begin{center}
    \includegraphics[width=0.45\textwidth, trim=0mm 0mm 0mm 0mm, clip=yes]{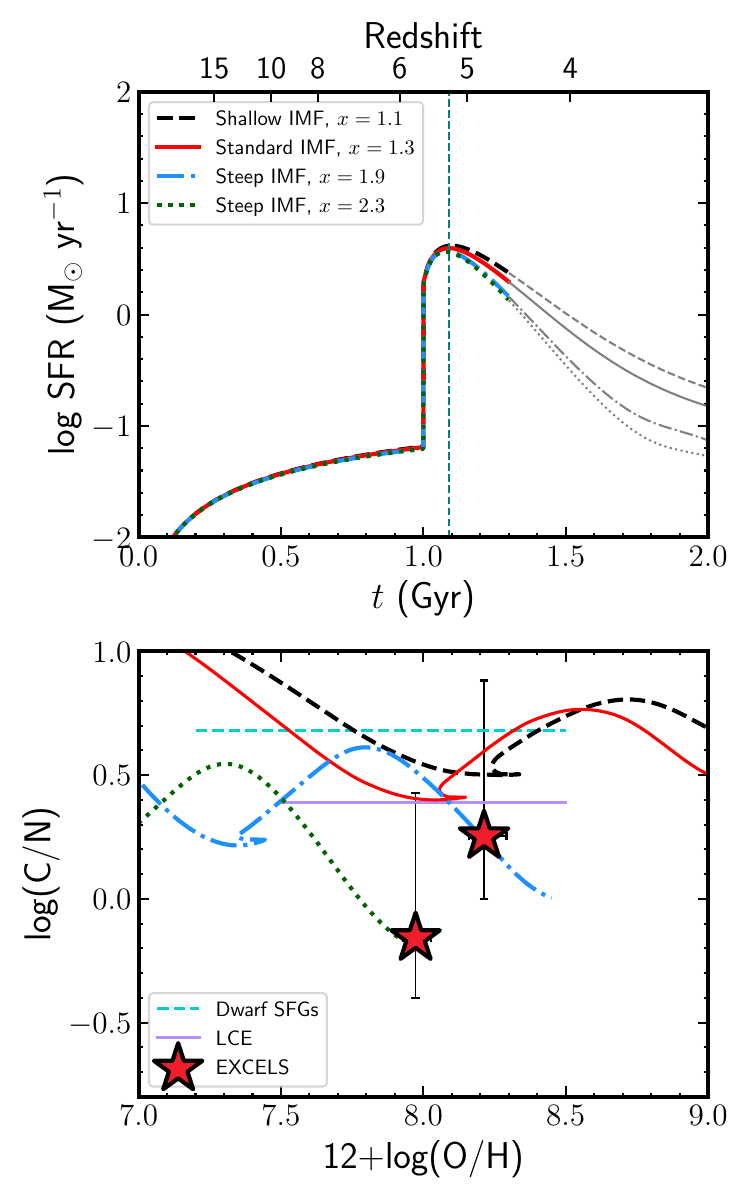}
        \caption{Chemical evolution models illustrating the affect of different IMFs on the $\rm C/N$ abundance ratios.
        {\textit{Top panel:}} A set of representative star-formation histories assuming a shallow IMF ($x=1.1$; black dashed line), a \citet{kroupa01} IMF ($x=1.3$; red solid line), and a steeper IMF ($x=1.9$ and $x=2.3$; dotted-dashed and dotted lines, respectively). 
        For reference, the teal dashed line shows the redshift of the EXCELS galaxies. 
        {\textit{Bottom panel:}} The evolution of the $\rm C/N$ ratio with respect to metallicity, compared with observations the EXCELS observations. 
        The chemical evolution models are generated based on the prescriptions of \citet{kobayashi20} and \citet{kobayashi24}.  
        The different lines indicate the IMF slope used in each model. 
        For reference, the cyan and purple lines indicate the mean value of C/N for local SFGs and LCE as in Fig.~\ref{fig:CN_relation}. 
        The chemical evolution models assuming bottom-heavy IMF are a better match to the C/N ratios of the EXCELS galaxies, although the uncertainties are clearly significant and a standard IMF is not ruled out.}
\label{fig:models}
\end{center}
\end{figure}

In the bottom panel of Fig.~\ref{fig:models}, we show the resulting chemical enrichment pathways in the $\rm{O/H}$ versus $\rm{C/N}$ plane and compare with our EXCELS observations. 
For reference, we have also indicated the mean values of $\rm{C/N}$ for the dwarf SFGs and LCE samples.
The characteristic up-and-down `U-shaped' model curves can be understood by considering the relative enrichment timescales of C and N.
After the onset of the second star formation episode, a galaxy will be rapidly enriched in O (increasing $\rm{O/H}$\footnote{The small decrease of O/H is caused by dilution with the gas infall that triggers the second star burst.} and decreasing (C,N)/O ratios) followed by N enrichment from intermediate-mass AGB stars ($M_{\odot} \simeq 4-7 \, \mathrm{M}_{\odot}$) that will cause a decrease in $\rm{C/N}$ (and increase N/O ratio).
Eventually, low-mass AGB stars ($\simeq 1 - 4 \, \mathrm{M}_{\odot}$) will begin to enrich the ISM with C after $t\simeq1.3$ Gyr, which is not shown in this figure.
As can be seen in Fig. \ref{fig:models}, for a steeper IMF galaxies will reach a lower value of $\rm{C/N}$ before C-enrichment from low-mass stars kicks in. The reason for this is because, for a steeper IMF, there are a greater proportion of N-enriching $M_{\odot} \simeq 4-7 \, \mathrm{M}_{\odot}$ stars relative to the massive stars $M_{\odot} \gtrsim 10 \, \mathrm{M}_{\odot}$ that provide the initial C-enrichment.

It can be seen that the standard IMF model is able to reproduce the mean values of $\rm{C/N}$ for the local samples (cyan and purple lines). 
Although this standard IMF model is also formally consistent with EXCELS-70864 and EXCELS-121806 within the uncertainties, it falls systematically above both observed $\rm{C/N}$ ratios.
Instead, we find that the steeper IMF models seem to align better with the $\rm{C/N}$ ratios observed in the EXCELS galaxies. 
Taken at face value, this comparison suggests that, if anything, a steeper IMF provides a better explanation for the observed low $\rm{C/N}$ ratios in metal-enhanced, evolved systems at $z\simeq5$.

This is different from what is found for the metal-poor galaxy at $z\simeq9$ presented in \citet{curti24b}. 
For that galaxy, the observed CNO abundances were explained by a single star burst (with much shorter timescales: $\tau_{\rm s}=0.0002, \tau_{\rm i}=0.001$ Gyr), and a top-heavy IMF for PopIII ($x=0$ for $30-120M_\odot$) was preferred.
Our galaxy is $0.6$ dex more metal-rich, and we find that it is not possible to explain the observed CNO abundances with a single star burst in this case.
For more evolved, metal-rich systems extended star-formation histories are needed to account for the build up in overall metallicity, and because of this AGB enrichment from intermediate mass stars ($M_{\odot} \simeq 4-7 \, \mathrm{M}_{\odot}$) also needs to be considered.
The simultaneous need to a top- and bottom-heavy to explain $\rm{C/N}$ ratios at $z>5$ across a wide range of metallicities might be achieved using, for example, a concordance IMF \citep[e.g.][]{vandokkum2024}.

However, for now the comparison is purely illustrative. 
Much larger samples of C and N measurements and more sophisticated modeling (i.e., complex star-formation histories, the inclusion of stellar feedback mechanisms) are clearly needed before any firm conclusion can be drawn.
Nevertheless, these models demonstrate that, contrary to metal-poor systems, low $\rm{C/N}$ ratios in more metal rich systems are less likely to be a result of a top-heavy IMF.

\section{Summary and Conclusions}
\label{sec:conclusion}

We have presented a study of the physical properties and chemical composition of two galaxies at $z\simeq5$ from the JWST-EXCELS survey \citep{carnall24}. 
Both galaxies benefit from medium resolution ($\mathrm{R}=1000$) JWST/NIRSpec spectroscopy covering their rest-frame UV to optical spectra across the wavelength range ${\lambda_{\rm rest} \simeq 1600-8000}$\AA.
This wide rest-frame wavelength coverage enables the measurement of several key nebular emission-lines.
Crucially, via the detection of the faint [\oiii]~\W4363 auroral line in both objects, we have derived robust chemical abundance estimates for several elements.
In this paper, we have focused in particular on the $\mathrm{C/O}$ and $\mathrm{N/O}$ abundance ratios which are accessible due to the detection of the UV \ciii]~\W1909 and optical [\nii]~\W6584 features. 
We have compiled a literature sample of star-forming galaxies at $z\simeq0$ up to $z\simeq10$ with which we can compare our results.
Crucially, there are currently only four other examples of galaxies at $z \gtrsim 5$ with direct estimates of C, N, and O. 
Our measurements therefore significantly expands upon the current sample.
Finally, we have interpreted our measured abundance ratios using the chemical evolution models of \citet{kobayashi20}.
Our main results can be summarised as follows:

\begin{itemize}

  \item Both galaxies are relatively metal-rich for their stellar mass (Fig.~\ref{fig: SFR_comparison}).
  For EXCELS-70864 we estimate ${\mathrm{log}(M_{\star}/\mathrm{M}_{\odot}) = {8.09}^{+0.24}_{-0.15}}$ and $\mathrm{12+log(O/H)} = 8.21^{+0.08}_{-0.05}$ ($Z \simeq 0.3 \, \mathrm{Z}_{\odot}$) while for EXCELS-121806 we estimate ${\mathrm{log}(M_{\star}/\mathrm{M}_{\odot}) = {8.02}^{+0.06}_{-0.08}}$ and $\mathrm{12+log(O/H)} = 7.97^{+0.05}_{-0.04}$ ($Z \simeq 0.2 \, \mathrm{Z}_{\odot}$).
  Interestingly, both galaxies appear to lie within the high-metallicity tail of the $z \gtrsim 4$ MZR and are fully consistent with the average MZR measured for the local ($z\simeq0$) high-redshift analogue sample from the CLASSY survey (Fig.~\ref{fig: SFR_comparison}).

  \item We measure $\mathrm{H}\alpha$-based star-formation rates of $6.2 \pm 0.1 \, \mathrm{M}_{\odot}\mathrm{yr}^{-1}$ and $10 \pm 0.2 \, \mathrm{M}_{\odot}\mathrm{yr}^{-1}$ for EXCELS-70864 and EXCELS-121806, respectively, meaning that both galaxies are consistent with the $M_{\star} - \mathrm{SFR}$ sequence formed from the local analogue and high-redshift ancillary samples (Fig.~\ref{fig: SFR_comparison}).
  We therefore find a striking similarity between our two $z\simeq5$ galaxies and local dwarf star-forming galaxies from CLASSY in terms of their global physical properties, namely stellar mass, SFR, and total metallicity.

  \item However, a detailed inspection of their chemical abundance patterns reveals some differences. 
  Focusing first on the $\mathrm{N/O}$ abundance ratio, we find that EXCELS-70864 and EXCELS-121806 fall within the high-$\mathrm{N/O}$ tail of the local distribution (Fig.~\ref{fig:CNO_excels}).
  Unlike other notably extreme examples of high $\mathrm{N/O}$ at $z\gtrsim6$ in the literature \citep[e.g.][]{Bunker23, topping24a} the $\mathrm{N/O}$ values we measure can be found among local systems and are therefore not unprecedented.
  However, our measurements continue a trend in which the majority of direct $\mathrm{N/O}$ measurements at $z \gtrsim 5$ fall systematically above the average local $\mathrm{N/O}-\mathrm{O/H}$ relation, with $\mathrm{log(N/O)} \gtrsim -1.0$ (Fig.~\ref{fig:CNO_excels}).
  These results might suggest the, even at higher metallicities, a source of N-enrichment exists that is common in the early Universe is absent in typical local dwarf star-forming galaxies.

  \item In contrast to the relatively high $\mathrm{N/O}$ ratios, we find that the $\mathrm{C/O}$ ratios are low and fall below the local $\mathrm{C/O}-\mathrm{O/H}$ relation (Fig.~\ref{fig:CNO_excels}).
  In general, we find that the $\mathrm{C/O}$ ratios we measure are consistent with other $z\gtrsim5$ galaxies in the literature which all fall close to the pure CCSNe yield expectation of $\mathrm{log(C/O)} \simeq -1.1$.
  These results suggest the C is enriched mainly by CCSNe at these high-redshifts, without significant input from low-mass AGB stars \citep{kobayashi20}.

  \item Combing the C, N and O abundances in the $\mathrm{C/N}$ versus $\mathrm{O/H}$ plane reveals a separation between the local and high-redshift galaxies (Fig.~\ref{fig:CNO_excels}).
  While all local galaxies lie at $\mathrm{log(C/N)} > 0$, the vast majority of high-redshift systems fall below this threshold.
  Again, these results suggest a preferential enrichment of N over C in the ISM of galaxies at $z \gtrsim 5$.
  Our new results suggest this the excess in N relative to C might also extend to moderately enriched galaxies ($Z \simeq 0.2-0.3 \, \mathrm{Z}_{\odot}$; Fig.~\ref{fig:CNO_excels}).

  \item To gain further insight into the chemical abundance ratios of our $z\simeq5$ galaxies, we explore a variety of scenarios using the chemical evolution models of \citet{kobayashi20}.
  We model the star-formation histories as steadily rising followed by a recent burst which elevates the SFR to $\simeq 10 \, \mathrm{M}_{\odot}\mathrm{yr}^{-1}$.
  In general, we find that models with steeper IMF slopes (i.e. a bottom-heavy IMF) do a better job at explaining the observed abundance ratios (Fig.~\ref{fig:models}).
  The fundamental reason for this is that a steeper, more bottom-heavy IMF provides more intermediate mass AGB stars ($M_{\odot} \simeq 4-7 \, \mathrm{M}_{\odot}$) to enrich the ISM with primary production of N.
However, given the uncertainties, our analysis is also fully consistent with standard IMF models.
  Ultimately, more data are needed to draw robust conclusions, but our models demonstrate that, contrary to the most metal-poor systems \citep[e.g.][]{curti24b}, a more top-heavy IMF is likely not favoured as an explanation for low $\rm C/N$ ratios in moderately-enriched high redshift galaxies.

 \end{itemize}
Our analysis shows that deep JWST data can provide robust, direct-method, chemical abundance ratios for star-forming galaxies at $z \simeq 5$.
We find that the preferential enrichment of N seen in low metallicity galaxies at high-redshift might also extend to more moderate metallicities ($Z \simeq 0.2-0.3 \, \mathrm{Z}_{\odot}$).
Contrary to the top-heavy IMF invoked to explain low $\rm{C/N}$ ratios in metal-poor high redshift galaxies, a comparison with chemical evolution models suggests that, if anything, a steeper, bottom-heavy IMF is favoured by our data.
However, the current sample sizes remain small, and more data is needed to any draw firm conclusions.
Thankfully, the wavelength coverage of NIRSpec means that similar studies can be carried out across the redshift range $4 < z < 7$; dedicated programs should therefore capable of significantly increasing sample sizes of galaxies with direct C, N and O abundances.

\section*{Acknowledgments}

    K. Z. Arellano-C\'ordova, F. Cullen, T. M. Stanton, and D. Scholte acknowledge support from a UKRI Frontier Research Guarantee Grant (PI Cullen; grant reference: EP/X021025/1). A. C. Carnall acknowledges support from a UKRI Frontier Research Guarantee Grant [grant reference EP/Y037065/1].  Based on observations with the NASA/ESA/CSA James Webb Space Telescope obtained at the Space Telescope
Science Institute, which is operated by the Association of Universities for Research in Astronomy, Incorporated, under NASA contract NAS5-
03127. Support for Program number JWST-GO-03543.014 was
provided through a grant from the STScI under NASA contract NAS5-
03127. 
CK acknowledges funding from the UK Science and Technology Facilities Council through grant ST/Y001443/1.

    Software: \texttt{jupyter} \citep{kluyver16}, \texttt{astropy} \citep{astropy:2018, astropy:2022}, {\tt PyNeb} \citep{luridiana15}, {\tt matplotlib} \citep{matplotlib},
{\tt numpy} \citep{numpy}, {\tt scipy} \citep{scipy}.

For the purpose of open access, the author has applied a Creative Commons Attribution (CC BY) licence to any Author Accepted Manuscript version arising from this submission.
\section*{Data Availability}

    All data will be shared by the corresponding author upon reasonable request.


\bibliographystyle{mnras}
\bibliography{refs} 



\appendix


\bsp	
\label{lastpage}
\end{document}